\PassOptionsToPackage{hyphens}{url}      % allow long URLs to break at hyphens
\PassOptionsToPackage{dvipsnames}{xcolor} % extra named colors if xcolor loads

\documentclass[acmsmall, screen]{acmart}

\setcopyright{none}
\renewcommand\footnotetextcopyrightpermission[1]{}
\authorsaddresses{}
\usepackage{comment}
\usepackage{afterpage}
\usepackage{tabularx}
\usepackage{multirow}
\usepackage{arydshln}     % dashed table rules (load after booktabs/array)
\usepackage{enumerate}
\usepackage{enumitem}
\usepackage{subcaption}
\usepackage{pifont}
\usepackage{xspace}
\usepackage{makecell}
\usepackage{framed}
\usepackage[normalem]{ulem}   % \sout for \jdel
\usepackage{tikz}
\usepackage[most]{tcolorbox}
\usepackage{wrapfig}
\usepackage{cleveref}     % must load after hyperref (acmart already loaded it)

\crefname{section}{\S}{\SS}
\crefformat{section}{\S#2#1#3}     % see cleveref manual, section 8.2.1
\crefformat{subsection}{\S#2#1#3}
\crefformat{subsubsection}{\S#2#1#3}

\setlist{itemsep=0pt,parsep=0pt,topsep=0pt}  % more compact lists

\definecolor{dkgreen}{rgb}{0,0.6,0}
\definecolor{gray}{rgb}{0.5,0.5,0.5}
\definecolor{mauve}{rgb}{0.58,0,0.82}
\definecolor{codegreen}{rgb}{0,0.6,0}
\definecolor{codegray}{rgb}{0.5,0.5,0.5}
\definecolor{codepurple}{rgb}{0.58,0,0.82}
\definecolor{backcolour}{rgb}{0.95,0.95,0.92}
\definecolor{light-gray}{gray}{0.95}
\definecolor{americanrose}{rgb}{1.0, 0.01, 0.24}
\definecolor{darkGreen}{RGB}{34, 139, 34}

\hypersetup{colorlinks=true,
            linkcolor=blue,
            citecolor=blue,
            urlcolor=blue}

\newtcolorbox{observation}{
    center,
    width=\linewidth,
    colframe=light-gray,
    colback=light-gray
}

\newtcolorbox{takeaway}{
    enhanced,
    width=\linewidth,
    colback=light-gray!50,
    colframe=black!75,
    boxrule=0.4mm,
    arc=2.5mm,
    fontupper=\fontsize{11pt}{13.5pt}\selectfont\itshape,
    left=2mm, right=2mm, top=1mm, bottom=1mm,
}

\title{To Store or To Regenerate? A Cost Model for AI-Generated Content at Scale}
\author{Yunjia Zheng}
\affiliation{
  \institution{Harvard University}
  \country{USA}
}

\author{Zirui Wang}
\affiliation{
  \institution{University of Virginia}
  \country{USA}
}

\author{Haoran Ni}
\affiliation{\institution{Harvard University}\country{USA}}

\author{Tingfeng Lan}
\affiliation{\institution{University of Virginia}\country{USA}}

\author{Zhaoyuan Su}
\affiliation{\institution{University of Virginia}\country{USA}}

\author{Yue Cheng}
\affiliation{\institution{University of Virginia}\country{USA}}

\author{Juncheng Yang}
\affiliation{\institution{Harvard University}\country{USA}}

\begin{document}

\begin{abstract}
% \yuec{need to be careful here, as HBM/DRAM price hikes, GPU prices increase, too}
AI-generated content is becoming a rapidly growing class of digital artifacts. Because these artifacts accumulate over time, their exponential growth creates a substantial storage, energy, and infrastructure cost for operators and society. At the same time, GPU compute cost continues to fall rapidly with each hardware generation. This divergence raises a fundamental question: when does on-demand regeneration become cheaper than persistent storage?

This paper develops a cost model for comparing persistent storage and on-demand regeneration for AI-generated artifacts. The model accounts for corpus growth, HDD and tape price trends, drive replacement, electricity, request skew, caching, generator FLOPs, and future GPU price-performance improvements. For image generation, our analysis shows that prompt-based regeneration does not become cheaper than storage until around 2040, because every cache miss must still rerun the full prompt-to-artifact generation pipeline.

We observe that widely used diffusion-based generation models operate in latent space, creating an alternative point in the cost tradeoff: instead of storing the final artifact or only the prompt, operators can store a compact intermediate representation (IR) and perform cheap on-demand decoding. Our analysis shows that caching combined with IR-based regeneration substantially reduces both storage and compute cost, making it at least 2$\times$  cheaper than both full-object storage and prompt-based regeneration even today. On a production image trace with 2.07 billion requests, the same conclusion holds: prompt-based regeneration is over 100$\times$ more expensive than storage, while IR-based regeneration reduces total cost to roughly half that of full-object storage while preserving interactive miss latency.

\end{abstract}

\maketitle

%sections
\section{Introduction}
\label{sec:introduction}

AI-generated content (AIGC) has rapidly become one of the largest and fastest-growing sources of digital artifacts. Unlike traditional digital content, which is typically captured, authored, or manually designed, AIGC artifacts are synthesized from compact user intent: a prompt, a reference image, or another lightweight specification. Users can generate new artifacts from scratch, edit existing ones, or regenerate them with small changes. Within a single year of widely used text-to-image services~\cite{ramesh2022dalle2,midjourney2022,rombach2022high}, users generated roughly as many images as photographers had captured during the first 150 years of photography~\cite{everypixel2024aistats}. This growth is not only large in absolute scale, but also accelerating as generation becomes integrated into mainstream applications. When ChatGPT introduced native image generation in 2025, more than 130 million users produced over 700 million images in the first week alone, creating enough demand that OpenAI reported its GPUs were ``melting'' and imposed emergency rate limits~\cite{openai2025gpumelting}. Similar growth pressures now appear beyond images, and data-center electricity consumed to generate and serve these artifacts is also rising steeply~\cite{iea2025energyai}.
% As AIGC corpora continue to expand across modalities, systems should not only store the growing generated artifacts, but also handle a faster pace of artifact creation and reuse.

This rapid expansion of AIGC turns storage into a growing systems burden. Because generated artifacts are cumulative, each newly created content adds to a long-lived corpus that must remain available for future retrieval, reuse, or regeneration. For HDD-based storage, the cost therefore extends far beyond the initial purchase of disks: drives must stay powered and cooled, electricity prices accumulate over time, and failed or aging drives must be replaced throughout the retention period. Moving colder artifacts to tape may appear to solve this problem because tape media is cheaper and consumes little power when idle. However, large-scale tape systems still require libraries, drives, robotics, maintenance, and operational management, and their high access latency makes them unsuitable for interactive serving. As a result, materialized-artifact storage is expensive because its costs accumulate across capacity, energy, replacement, and archival infrastructure.

At the same time, the cost of compute is falling fast. GPU \$/TFLOPS declines exponentially across past GPU generations: roughly 30\% per year for FP16 throughput (from 92\,\$/TFLOPS for V100 to 16\,\$/TFLOPS for B200), and about 41\% per year for lower-precision tensor computation (\autoref{sec:price_trend}). This divergence raises a question: 

\vspace{2pt}
\textit{For a specific duration, when is it cheaper to regenerate on demand than to store?}
\vspace{2pt}

Answering this break-even question is not a simple comparison between today’s storage price and generation cost. It requires modeling two evolving serving pipelines. For materialized-artifact storage, the cost depends on corpus growth, retention duration, HDD and tape price trends, electricity prices, replacement cycles, and maintenance of archival infrastructure. For regeneration, the cost depends not only on the generator’s FLOPs and future GPU price-performance trends, but also on how artifacts are accessed. By exploiting the access patterns, a system can improve the efficiency and reduce the cost with proper caching strategies. Finally, generation cost itself is not fixed: model efficiency, GPU pricing, and low-precision acceleration can all shift the break-even year. We therefore build a cost model that jointly captures materialized-artifact storage, regeneration, caching, and hardware trends to determine when generated artifacts should be retained in materialized form and when they should instead be regenerated on demand.
\begin{figure}
  \centering
  \begin{subfigure}{0.49\textwidth}
    \centering
    \includegraphics[width=\textwidth]{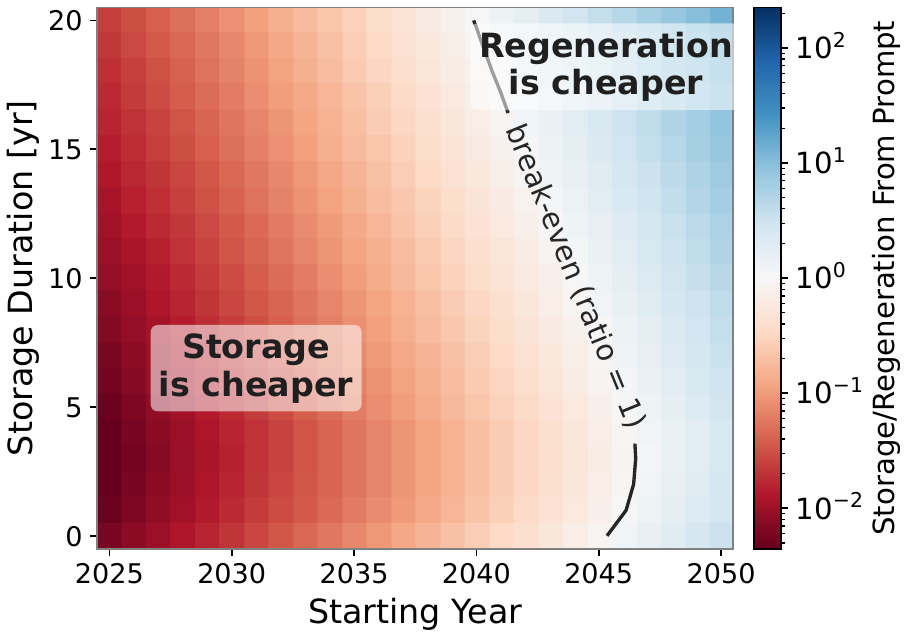}
    \caption{Materialized-artifact storage is cheaper than prompt-based regeneration most of the time.}
    \label{fig:startyear-fullregen}
  \end{subfigure}\hfill%
  \begin{subfigure}{0.48\textwidth}
    \centering
    \includegraphics[width=\textwidth]{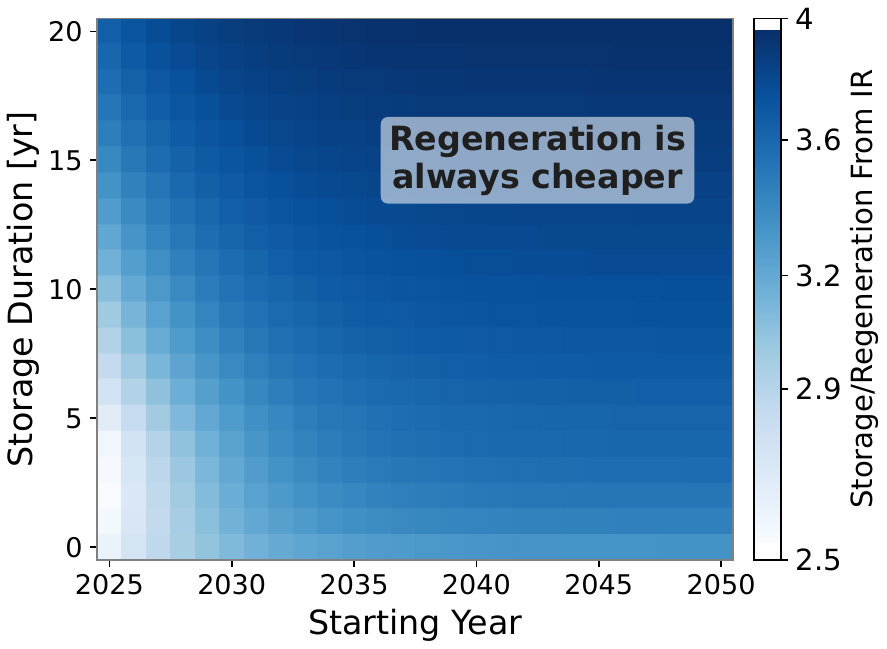}
    \caption{Caching and storing intermediate representation (LazIR) make regeneration cheaper.}
    \label{fig:startyear-twotier}
  \end{subfigure}
  \vspace{-5pt}
  \caption{When will regeneration be cheaper than materialized-artifact storage for AI-generated artifacts? Using image generation as an example, the figure shows the cost ratio of materialized-artifact storage on HDD to regeneration. Values above 1 mean regeneration is cheaper than materialized-artifact storage. (a) With prompt-based regeneration, materialized-artifact storage remains cheaper in the near future. As GPU costs continue to fall, prompt-based regeneration eventually becomes cheaper from 2046 if artifacts are stored for 5 years. (b) Regenerating from a stored intermediate representation (IR) lowers the regeneration cost substantially. As a result, IR-based regeneration is already cheaper than materialized-artifact storage starting today, regardless of the retention duration.}
  \vspace{-10pt}
  \label{fig:startyear-image}
\end{figure}

Unsurprisingly, our cost model shows that prompt-based regeneration is significantly more expensive than materialized-artifact storage. 
The key observation is that AI-generated artifacts are not produced in a single step, but in a multi-stage pipeline. A generator typically transforms a compact input through several intermediate stages, such as prompt conditioning, iterative sampling, latent refinement, and final decoding into various artifact formats.
Instead of storing only the prompt or the materialized artifact, we can store a compact \textit{intermediate representation} (IR), for example the latent for text-to-image generation~\cite{ho2020ddpm,saharia2022imagen,rombach2022ldm}, that captures the costly part of generation after it has already been computed. Regenerating from this latent creates a fast path: the system skips the most expensive upstream stages and only runs the lightweight final decoding step, reducing storage cost without paying the full cost of prompt-based regeneration.

Figure~\ref{fig:startyear-image} previews the crossover between regeneration and storage for image generation under exponential corpus growth. With prompt-based regeneration, the crossover occurs only after 2040: regeneration is initially too expensive because every request reruns the full generator, and although caching avoids regeneration for hot objects, every cache miss still invokes the same costly pipeline. In contrast, the crossover arrives immediately when regeneration starts from an intermediate representation. By caching hot objects, storing the cold tail as compact latents, and regenerating through a cheaper decoding path, the IR-based design makes regeneration cheaper by roughly $2.6\times$ to $4.0\times$. In other words, \textit{choosing IR as the retained state shifts the break-even point from the future to today no matter how long the data is stored}, making on-demand regeneration economically favorable from the start.

This paper makes the following contributions:
\begin{itemize}[noitemsep,leftmargin=*]
\item An \textbf{empirical characterization of AIGC growth and hardware trends}. We collect and fit historical trends for generated-corpus growth, HDD and tape storage cost, HDD power density, GPU \$/TFLOPS, and GPU energy per TFLOP, showing that compute improves substantially faster than storage.

\item A \textbf{closed-form lifecycle cost model for storage and regeneration}. The model covers materialized-artifact storage on HDD and tape, accounting for corpus growth, retention duration, price trends, infrastructure overhead, replacement, and electricity, as well as regeneration policies including prompt-based regeneration, hot-set caching under skewed access, and storing with \textbf{LazIR}, an IR store with cache. It provides an explicit break-even threshold for deciding when regeneration becomes cheaper than materialized-artifact storage.

\item A \textbf{quantitative evaluation across image and music generation workloads}. 
% We show that materialized-artifact storage is 100$\times$ cheaper than prompt-based regeneration. 
% However, storing intermediate representation (IR) consistently beats materialized-artifact storage across modalities and under both linear and exponential data growth.
We show that prompt-based regeneration is prohibitively expensive, with materialized-artifact storage remaining nearly $100\times$ cheaper. In contrast, LazIR consistently outperforms materialized-artifact storage, reducing lifecycle cost by $3.1\times$–$3.7\times$ for image workloads and $8.6\times$–$10.5\times$ for music workloads under linear and exponential corpus growth.

\item An \textbf{evaluation using production trace}. We confirm our finding using a trace replay of a 2.87-year-long production trace from a leading image generation platform---LazIR reduces cost by around 50\% compared to storing materialized artifacts.  
\end{itemize}

\section{Generated-Data Growth and Computing Cost Trends}
\label{sec:cost_trends}

\subsection{Growth of AI-Generated Data}
Generative models have become a first-class source of new data, with outputs accumulating at an exponential pace across modalities. Adobe Firefly, for example, grew from roughly 70 million generated images at launch in mid-2023 to about 6.5 billion within one year and more than 22 billion by early 2025~\cite{adobeimage}.
Similar growth has been reported by other providers. For example, an estimated 15 billion AI-generated images were produced between 2022 and 2023, while daily AI-image generation requests increased from a few million per day to roughly 80 million per day by 2026~\cite{everypixel2024aistats,aiimagevolume2026}. For comparison, traditional photography took 149 years to reach 15 billion images~\cite{everypixel2024aistats}.

% These are the outputs of single services in each modality, yet they already
% imply tens of billions of objects that an operator may be expected to retain indefinitely.
% The convex shape motivates the growing-corpus model of
% ~\autoref{sec:growing-corpus}, where new data arrives at a compounding
% rate of $g$ per year rather than at a fixed annual volume.

% \subsection{Electricity demand}
% \jason{why do we need this?}
% The electricity drawn by compute and storage is rising steeply over time.
% The Base Case of the IEA \emph{Energy and AI} report~\cite{iea2025energyai}
% projects global data-centre electricity demand.
% Electricity demand more than doubles from around 270\,TWh in 2020 to about
% 415\,TWh in 2024 and around 945\,TWh by 2030, a compound growth of roughly
% 12 to 15\% per year that is driven mainly by accelerated computing for AI.
% % The associated CO$_2$ emissions climb from around 180\,Mt today toward a
% % peak near 320\,Mt by 2030.
% % Emissions grow more slowly than electricity because the grid keeps
% % decarbonising, yet the absolute tonnes of CO$_2$ from compute and storage
% % still rise steeply.
% The message is that the energy footprint of data centres is on a
% steep upward trajectory, so it cannot be treated as a negligible externality
% and instead must be priced into the storage versus recompute decision.

\subsection{Storage and Compute Cost Trends}
\label{sec:price_trend}

\noindent\textbf{HDD \$/TB.}
% \jason{we may show a density figure as well especially we can have 40 years of trend}
Consumer HDD price per terabyte declined by around 9\% annually
from 2010 (\$46/TB) to 2022 (\$14.40/TB), according to a
monthly survey~\cite{mccallum2023diskprice}.
With Heat-Assisted Magnetic Recording (HAMR) re-accelerating
areal-density growth to around 20\%/year~\cite{wikipedia_hdd},
we conservatively model a \textbf{10\%/year} decline in \$/TB going
forward, anchored at \$20/TB in 2025~\cite{diskprices2025}.
% $C_{\text{HDD}}(t)=20\times 0.90^{(t-2025)}$ [\$/TB].
% \jason{we may need to mention that each new technique brings a step change---we will not see 20\% continue}

\begin{figure*}[t]
  \centering
  % \begin{subfigure}{0.48\textwidth}
  %   \centering
  %   \includegraphics[width=\textwidth]{figures/modality_growth.pdf}
  %   \caption{Cumulative generated content.}
  %   \label{fig:imggen}
  % \end{subfigure}\hfill
  \begin{subfigure}{0.32\textwidth}
    \centering
    \includegraphics[width=\textwidth]{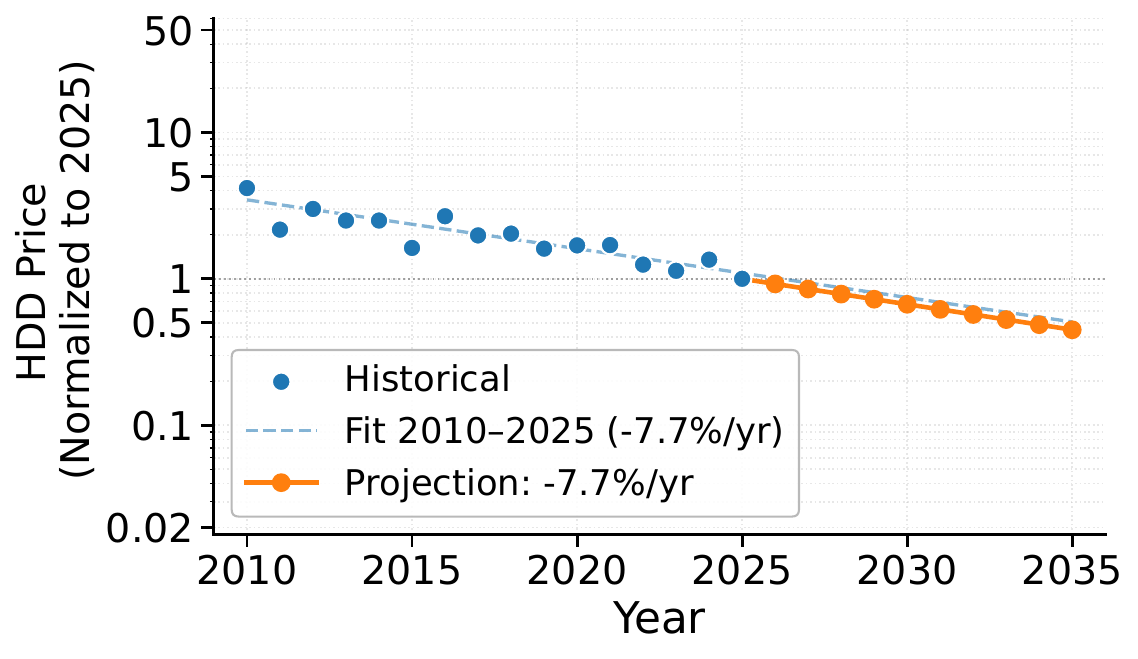}
    \caption{Normalized HDD cost.}
    \label{fig:hdd-price}
  \end{subfigure}
  % \\[4pt]
  \begin{subfigure}{0.32\textwidth}
    \centering
    \includegraphics[width=\textwidth]{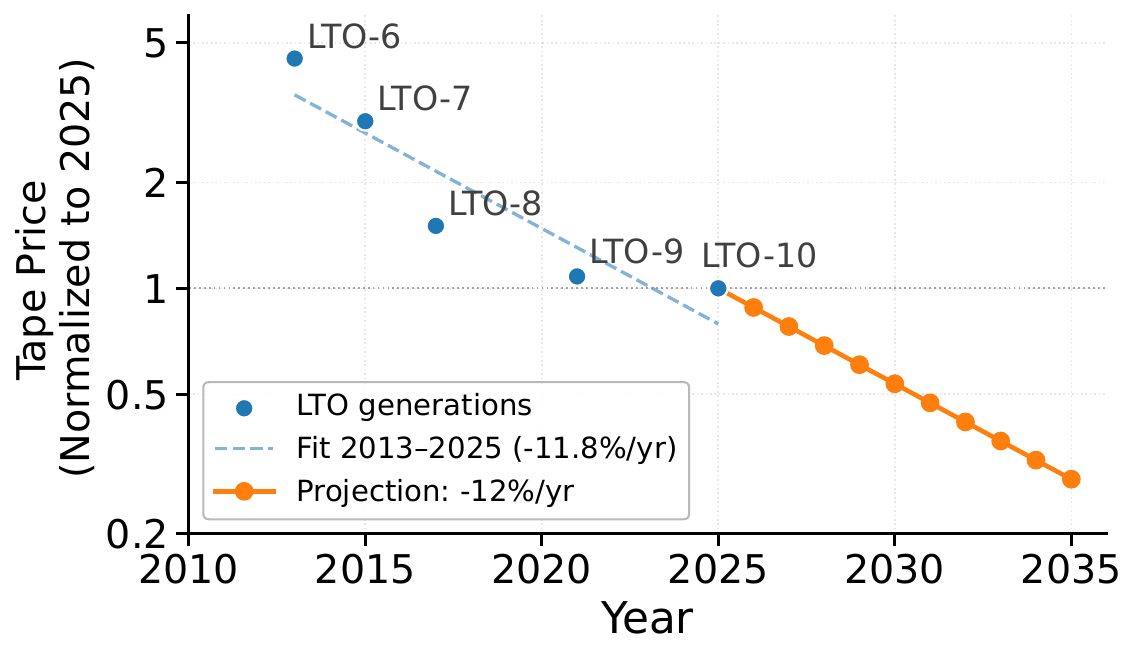}
    \caption{Normalized Tape cost.}
    \label{fig:tape-price}
  \end{subfigure}\hfill
  \begin{subfigure}{0.32\textwidth}
    \centering
    \includegraphics[width=\textwidth]{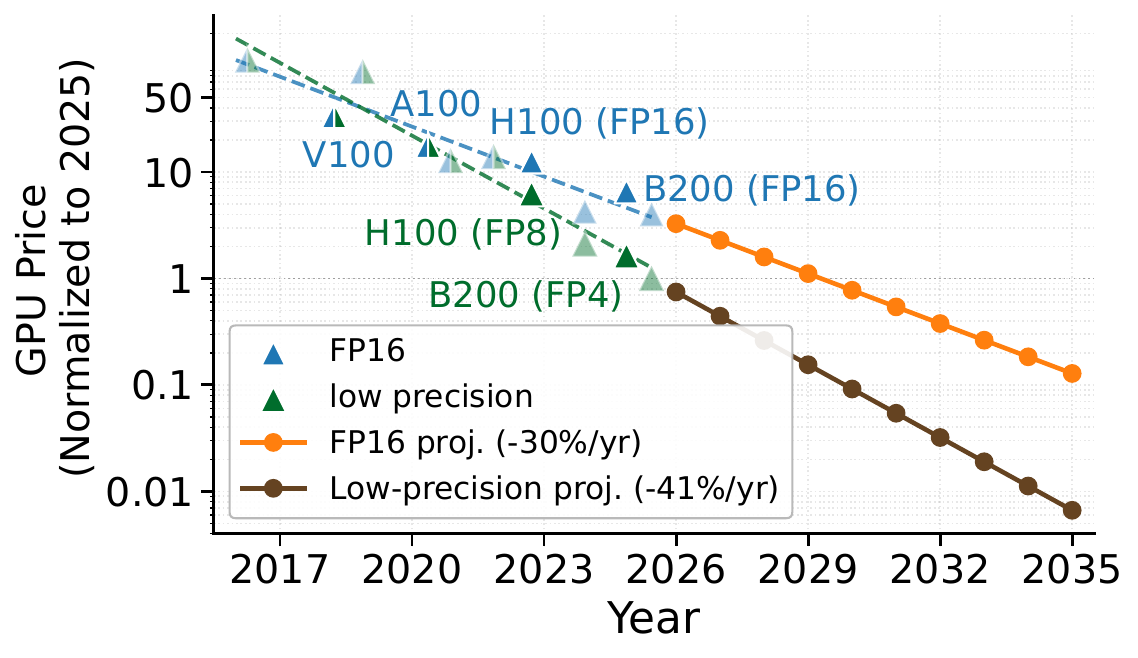}
    \caption{Normalized Cost per TFLOPS.}
    \label{fig:gpu-price-perf}
  \end{subfigure}
  \vspace{-5pt}
  \caption{Storage and compute costs decline at different rates. HDD and tape prices fall by roughly 10\% per year, while GPU \$/TFLOPS drops much faster, at about 30\% per year for FP16 throughput and 41\% for low-precision computation (all sources listed in \autoref{sec:appendix_source} in the appendix).~\cite{mccallum2023diskprice,epochAI,tapeandmedia_ltoprices,techradar2025lto10}.}
  \label{fig:price-trends}
\end{figure*}

\noindent\textbf{Tape \$/TB.}
Tape remains the cheapest archival tier, but its price-per-terabyte is
declining more slowly than HDD and appears to be flattening.
Native \$/TB across LTO generations falls from \$46.9/TB for LTO-6 in 2013 to
\$15.7/TB for LTO-8 in 2017 and \$11.3/TB for LTO-9 in 2021, yet LTO-10 in 2025
landed at only \$10.4/TB, barely below its predecessor despite doubling
capacity~\cite{tapeandmedia_ltoprices,techradar2025lto10}.
The early generations imply a roughly 12\%/year decline, but the LTO-9 to
LTO-10 step is nearly flat, so we treat tape as a slowly improving floor rather
than a source of continued exponential savings.
\autoref{fig:tape-price} shows this trend normalized to 2025.

\noindent\textbf{GPU cost at dollar per TFLOPS for FP16.}
\autoref{fig:gpu-price-perf} plots data-center GPU price performance from
the Epoch AI machine-learning hardware dataset~\cite{epochai2024mlhardware}
and overlays two series.
The first prices every GPU at its dense FP16 rate, the matrix-engine
throughput that governs modern ML inference, and an exponential fit yields a \textbf{30\%/year} decline in
FP16 \$/TFLOPS, a halving of price per unit throughput roughly every two
years.
The second instead prices each GPU at the lowest numeric precision its
tensor cores support natively, whose \$/TFLOPS is several times lower
than the FP16 rate.
Because each new low-precision format resets price-performance sharply
downward, this low-bit curve is steeper, declining by roughly
\textbf{41\%/year}, so inference served at reduced precision cheapens faster
than the FP16 trend alone implies.
We adopt the more conservative 30\%/year FP16 rate going forward, anchored at
\$18/TFLOPS in 2025, the dense FP16 price of an NVIDIA B200
(\$40{,}000 for 2{,}250\,TFLOPS).
% Capital is thus referenced to the same FP16 precision as the
% energy coefficients later in the discussion.

% \autoref{tab:facts} collects the exogenous, empirically grounded
% constants that drive the model, namely the hardware price trends,
% device lifetimes, and electricity price, which are fixed
% at the reported values.
%and not considered as parameters.
% \zirui{generation cost seems to be held fixed, but model compute demand keeps rising over time?}
% ~\autoref{tab:modality-params} lists the model- and workload-dependent
% parameters that we vary in the ablation analyses of
% \autoref{sec:growing-corpus}--\autoref{sec:recompute}.

% \noindent\textbf{Forward projection.}
% Combining the consumer-GPU fit ($\sim$29\%/yr) with the
% datacenter rate ($\sim$21\%/yr) and Epoch's aggregate figure
% ($\sim$30\%/yr), we adopt a \textbf{30\%/year} decline in
% \$/TFLOP from the 2025 anchor of \$19/TFLOP
% (NVIDIA RTX\,5090: 104.8 TFLOPS at \$1{,}999 MSRP):
% $c_{\text{GPU}}(t)=19\times 0.70^{(t-2025)}$ [\$/TFLOP].
% Under this model, consumer GPU cost falls to $\sim$\$3.2/TFLOP by 2030
% and $\sim$\$0.5/TFLOP by 2035, while HDD storage reaches
% $\sim$\$12/TB and $\sim$\$7/TB over the same horizons.

\begin{figure}[t]
  \centering
  \begin{subfigure}{0.4\columnwidth}
    \centering
    \includegraphics[width=\textwidth]{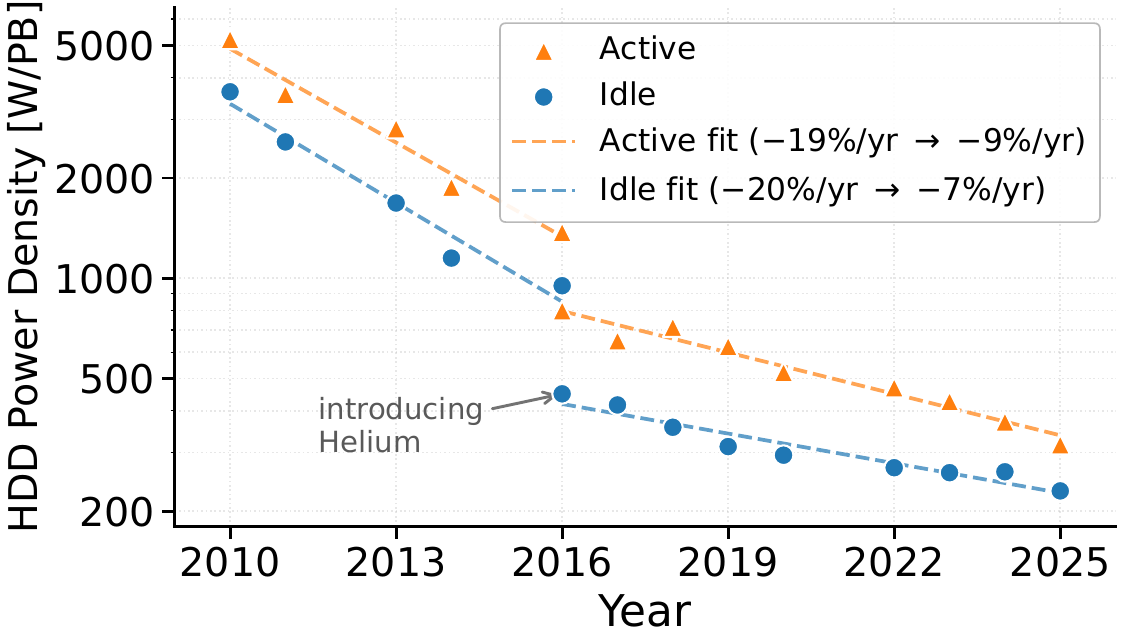}
    \caption{HDD power density [W/PB].}
    \label{fig:hdd-power}
  \end{subfigure}
  \hspace{2em}
  \begin{subfigure}{0.4\columnwidth}
    \centering
    \includegraphics[width=\textwidth]{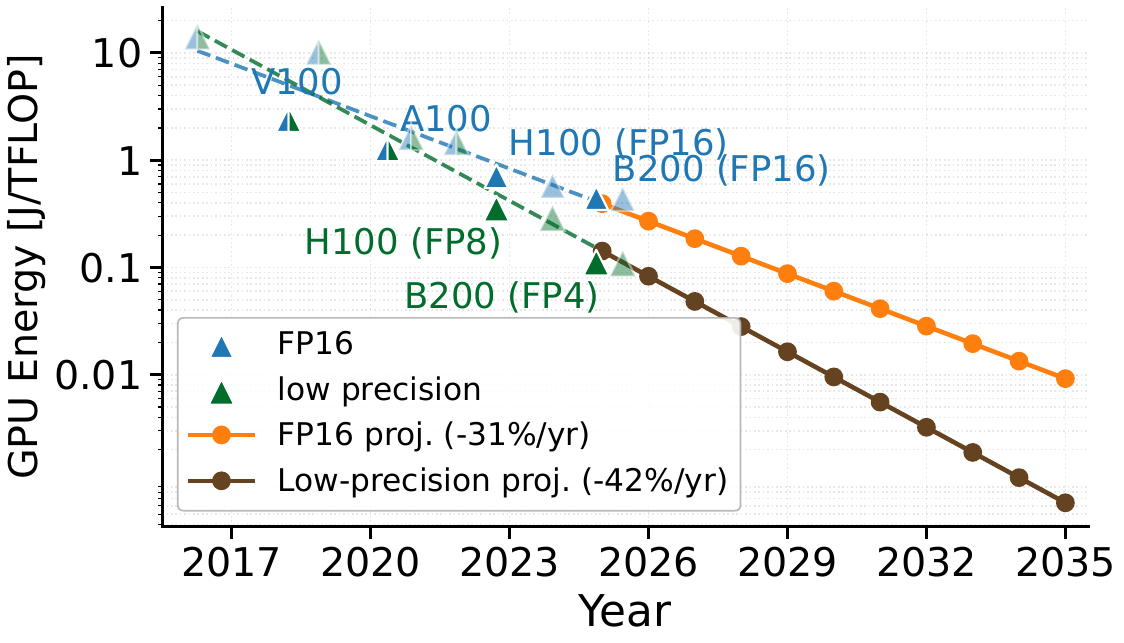}
    \caption{GPU energy per TFLOP [J/TFLOP].}
    \label{fig:gpu-energy}
  \end{subfigure}
  \vspace{-5pt}
  \caption{Energy required for unit of storage and compute declines at rates similar to respective price declines.} 
  \vspace{-10pt}
  \label{fig:power-trends}
\end{figure}

\subsection{Hardware Energy Consumption}
\label{sec:power-trends}

\autoref{fig:power-trends} tracks the energy efficiency of storage and computation over time, providing the empirical basis for the two unit costs for energy in our model.

\noindent\textbf{HDD power density.}
\autoref{fig:hdd-power} shows the idle power per stored petabyte for the Seagate enterprise nearline drive lineage from 2010 to 2025. The trend has two phases, separated by the introduction of Helium drives at 10 TB in 2016. Before helium, air-filled drives improved by roughly 20\% per year, falling from 3600 W/PB in 2010 to 950 W/PB in 2016, as drive capacity increased while per-drive power remained nearly constant. Helium technology then reduced power density further, roughly halving it to 450 W/PB. 
After this transition, however, the trend becomes much flatter, improving by only about 7\% per year and reaching 0.23 W/TB for the 30 TB Exos M. This slowdown occurs because helium drives continue to add platters and heads, causing per-drive power to rise with capacity and leaving W/TB nearly unchanged.

\noindent\textbf{GPU energy per TFLOP.}
\autoref{fig:gpu-energy} shows board TDP normalized by dense throughput for the same data-center accelerators used in the price-performance analysis. FP16 energy per TFLOP falls by about 31\% per year, from 14.14 J/TFLOP for the P100 in 2016 to 1.28 J/TFLOP for the A100 in 2020 and 0.44 J/TFLOP for the B200 in 2024. When measured at each GPU's lowest supported precision, energy per TFLOP falls even faster, at roughly 42\% per year.

Overall, the energy trends closely track the price trends used earlier: HDD power density improves at approximately the same rate as HDD cost per byte, while GPU energy per TFLOP declines at a rate comparable to GPU \$/TFLOPS. We therefore use these fitted energy-improvement rates in the cost model below, so that both monetary cost and electricity consumption evolve consistently across storage and regeneration pipelines.

% The model anchors the GPU energy coefficient at \textbf{1\,J/TFLOP} using the
% A100 data sheet, so the current fitted FP16 value near 0.4\,J/TFLOP is already
% about 2.6$\times$ lower, making the assumption conservative rather than optimistic.
% Both efficiency metrics improve over time. Our conservative anchors therefore
% understate, if anything, how inexpensive storage power and compute energy may
% become.

\section{Cost Model 1: Materialized-Artifact Storage}
\label{sec:hdd-cost-model}

We begin with the materialized-artifact storage baseline, in which every generated artifact is retained for the entire study horizon. We consider two ingest patterns: a constant 1\,PB per year and annual ingest that grows by a factor of $1+g$. The total cost includes initial storage purchases, HDD replacement, and the electricity required to keep the stored capacity powered. We evaluate both HDD and tape, whose different lifetimes and infrastructure requirements produce different cost structures. \autoref{tab:highlevel_notation} summarizes the notation conventions used throughout the cost models. 

\afterpage{%
\begin{table}[t]
  \centering
  \footnotesize
  \setlength{\tabcolsep}{5pt}
  \renewcommand{\arraystretch}{1.15}
  \caption{Notation conventions used in the cost models.}
  \label{tab:highlevel_notation}
  \begin{tabular}{@{}ll@{}}
    \toprule
    Symbol & Meaning \\
    \midrule
    $c_{\cdot}$ & Unit cost for storage, compute, or electricity. \\
    $C_{\cdot}$ & Monetary cost of storage, regeneration, caching, or total system operation. \\
    $S_{\cdot}$ & Data volume to store, cache, purchase, or keep live. \\
    $E_{\cdot}$ & Cumulative electricity cost over a modeled duration. \\
    $R_{\cdot}$ & Number of requests served over a period of time. \\
    \bottomrule
  \end{tabular}
\end{table}
}

\subsection{Parameter Anchors}
\label{sec:param-anchors}
\noindent\textbf{Workload assumptions.}
We evaluate the model using a running image-generation example with an object size of
% $s_{\text{obj}} = 3\times10^{-9}$\,PB,equivalently a 
3\,MB per image, an annual ingest growth rate of
$g = 20\%$, and an HDD service life $L_{\text{HDD}} = 5$\,years, over the
horizon 2025--2045.
% Every break-even cost we later report scales linearly with $s_{\text{obj}}$.

\noindent\textbf{Disk price.}
We instantiate the HDD price trend introduced in
\autoref{sec:cost_trends} with concrete anchors.
For storage, we take $C_{\text{HDD}}(2025) = 20{,}000$\,\$/PB~\cite{diskprices2025}
(equivalently 20\,\$/TB) and an annual decline
$\mathrm{drop}_{\text{HDD}} = 10\%$, so
$C_{\text{HDD}}(t) = 20{,}000\times 0.90^{(t-2025)}$\,[\$/PB].
The GPU price anchor, needed only once regeneration enters, is given in
\autoref{sec:regen-cost-model}.

\noindent\textbf{Electricity price.}
Keeping stored data powered is priced at an electricity rate of
$e = \$0.08$/kWh, a typical data-center tariff~\cite{eia_electricity}.
The unit cost of $c_{\text{elec,HDD}}$\,[\$/PB/yr] turns a medium's 2025 system power
density $P^{\text{sys}}$\,[W/PB] into the annual cost of keeping one petabyte
powered at the 2025 anchor,
\begin{equation}
  c_{\text{elec,HDD}}(2025) = P^{\text{sys}} \times 8{,}760\,\tfrac{\text{h}}{\text{yr}}
           \times \frac{e}{10^{3}},
  \label{eq:alpha}
\end{equation}
For HDD, the system power density is $P_{\text{HDD}}^{\text{sys}} = 450$\,W/PB,
combining the bare-drive active power of the 2025 Seagate Exos\,M 30\,TB
(9.5\,W, or about 320\,W/PB)~\cite{seagate_exosm_datasheet} with a PUE of 1.4 for
cooling and power-delivery overhead~\cite{uptime_pue2023}, so
$c_{\text{elec,HDD}}(2025) = 450 \times 8{,}760 \times 0.08/10^{3}
\approx 315$\,\$/PB/yr.
Rising areal density spreads a roughly fixed per-drive wattage over more
terabytes, so the HDD power density falls over time at about the same
$\mathrm{drop}_{\text{HDD}} = 10\%$/yr as the price, and the unit cost
declines from its 2025 anchor as
\begin{equation}
  c_{\text{elec,HDD}}(y) = c_{\text{elec,HDD}}(2025)\,(1-\mathrm{drop}_{\text{HDD}})^{(y-2025)},
  \quad c_{\text{elec,HDD}}(2025) \approx 315\;\text{\$/PB/yr}.
  \label{eq:alphahdd}
\end{equation}
The electricity rate $e$ itself is held at its 2025 value throughout; only the
device energy intensity declines.
Tape consumes far less power: idle cartridges draw no power, and only the mounted drives
and library controller consume electricity~\cite{fujifilm2024tco}. Rather than track a
separate tape power density, we model the unit cost of electricity for tape as a
fixed $10\%$ of that of HDD at every year, so it shares the HDD
$10\%$/yr decline and stays a constant $90\%$ below HDD,
\begin{equation}
  c_{\text{elec,tape}}(y) = 0.1\,c_{\text{elec,HDD}}(y),
  \quad c_{\text{elec,tape}}(2025) = 0.1 \times 315 \approx 31.5\;\text{\$/PB/yr}.
  \label{eq:alphatape}
\end{equation}
We use $c_{\text{elec,HDD}}$ for HDD in~\autoref{sec:power-cost} and
$c_{\text{elec,tape}}$ for tape in~\autoref{sec:tape-cost}. The electricity unit cost for GPU
compute $c_{\text{elec,GPU}}$\,[\$/TFLOP], needed only for regeneration, is
derived in~\autoref{sec:recompute}.

\subsection{Materialized-Artifact Storage on HDD}
\label{sec:hdd-store}

\subsubsection{Cumulative Capital Cost}
\label{sec:hdd-capital}

\noindent\textbf{Linear growth.}
\label{sec:steady-1pb}
We project the HDD capital cost of an accumulating corpus of
artifacts from 2025 to year $t$, assuming linear growth from a constant 1\,PB of
new data ingested every year, for which the HDD price declines geometrically from
its 2025 anchor at an annual rate $\mathrm{drop}_{\text{HDD}}$:
\begin{equation}
  C_{\text{HDD}}(y)
    = C_{\text{HDD}}(2025)\,(1-\mathrm{drop}_{\text{HDD}})^{(y-2025)}
    \;. \label{eq:phdd}
\end{equation}
The anchor $C_{\text{HDD}}(2025)$ and the decline rate
$\mathrm{drop}_{\text{HDD}}$ are instantiated in
\autoref{sec:recompute}.
Each year a 1\,PB tranche of HDD is purchased at that year's price.
The cumulative HDD capital cost through year $y$ is
\begin{align}
  C_{\text{HDD,cum}}(y)
    &= \sum_{\tau=2025}^{y} C_{\text{HDD}}(\tau).
  \label{eq:chddcum}
\end{align}

\noindent\textbf{Exponential growth.}
\label{sec:growing-corpus}
We now extend the model to exponential ingest: annual ingest grows at
rate $g$, so the new data ingested in year $\tau$ is
$S_{\text{data}}(\tau) = (1+g)^{\tau-2025}$\,PB, while the cumulative
corpus volume is the sum of all previous ingest. HDD is purchased
at price $C_{\text{HDD}}(\tau)$. The 1\,PB
baseline above is the special case $g=0$. The cumulative capital cost through year $y$ becomes
\begin{align}
  C_{\text{HDD,cum}}^{(g)}(y)
    &= \sum_{\tau=2025}^{y} S_{\text{data}}(\tau)\,C_{\text{HDD}}(\tau).
    \label{eq:chddcumg}
\end{align}

\subsubsection{HDD Replacement over the Corpus Lifetime}
\label{sec:hdd-replacement}

\autoref{eq:chddcum} and \autoref{eq:chddcumg} charge each petabyte
once, implicitly assuming a drive survives the full horizon.
Mechanical HDDs instead have a finite service life $L_{\text{HDD}}$,
after which the data they hold must be re-hosted on freshly purchased
drives.
We take $L_{\text{HDD}} = 5$\,years, a typical enterprise refresh
interval, so each retained petabyte is bought roughly three to four
times across 2025--2045.

\smallskip
\noindent\textbf{Annual provisioning with replacement.}
Because the corpus is cumulative, every petabyte ever ingested must
remain on live disks. Each year therefore purchases new-data capacity
plus a replacement for the cohort installed $L_{\text{HDD}}$ years
earlier:
\begin{equation}
  S_{\text{HDD,purchase}}(y) = S_{\text{data}}(y) + S_{\text{HDD,purchase}}(y-L_{\text{HDD}}),
  \label{eq:provision}
\end{equation}
with $S_{\text{HDD,purchase}}(y){=}0$ for $y<2025$, where the first term is freshly
generated data and the second re-buys the drives that reach
end-of-life.
Each replacement is priced at \emph{its own} year $C_{\text{HDD}}(\tau)$,
not the original purchase year, so replacement purchases benefit from the ongoing
price decline.

\smallskip
\noindent\textbf{Replacement-inclusive cumulative cost.}
Summing the annual purchases at their year-of-purchase price gives the
cumulative HDD cost used in all subsequent break-even analyses:
\begin{equation}
  C_{\text{HDD,cum}}^{(g)}(y)
    = \sum_{\tau=2025}^{y} S_{\text{HDD,purchase}}(\tau)\,C_{\text{HDD}}(\tau),
  \label{eq:chddcumrepl}
\end{equation}
with the linear-growth 1\,PB baseline recovered as its $g{=}0$ case,
$C_{\text{HDD,cum}}\equiv C_{\text{HDD,cum}}^{(0)}$ (where
$S_{\text{HDD,purchase}}(\tau)=1+S_{\text{HDD,purchase}}(\tau-L_{\text{HDD}})$).
The \autoref{eq:chddcum} and \autoref{eq:chddcumg}
are the $L_{\text{HDD}}\to\infty$ limits; henceforth $C_{\text{HDD,cum}}$
and $C_{\text{HDD,cum}}^{(g)}$ denote these replacement-inclusive costs.

\smallskip
\noindent\textbf{Live versus manufactured volume.}
Two volumes must be kept distinct.
The \emph{live} volume, the cumulative data $[(1+g)^{y-2024}-1]/g$, is
what stays powered and is unchanged by replacement, since dead drives
are swapped one-for-one.
The \emph{manufactured} volume $\sum_{\tau} S_{\text{HDD,purchase}}(\tau)$ drives
capital cost.

\subsubsection{Electricity Cost}
\label{sec:power-cost}

The analysis so far counts only capital expenditure on HDD hardware.
We now add the operational electricity cost of keeping every stored
petabyte powered at all times.

\smallskip
\noindent\textbf{Electricity unit cost.}
Operational electricity enters through one aggregate unit cost: unit cost for HDD $c_{\text{elec,HDD}}(t)$, anchored at
$c_{\text{elec,HDD}}(2025) = 315$\,\$/PB/yr and declining at $\mathrm{drop}_{\text{HDD}} =
10\%$/yr, the annual cost of keeping one petabyte of HDD powered, derived from
the HDD power density and the electricity rate in
\autoref{sec:param-anchors}, \autoref{eq:alphahdd}.
A petabyte provisioned in an earlier year is billed at the current year's
cost, since dead drives are replaced by newer, denser
power-efficient models.

\smallskip
\noindent\textbf{Annual HDD electricity.}
All cumulative HDD capacity from past vintages must remain powered.
The total powered HDD volume in year $y$ is the cumulative corpus, the running
sum of the annual ingest,
$S^{\text{full}}_{\text{data}}(y) = \sum_{k=2025}^{y} S_{\text{data}}(k)
= \sum_{k=2025}^{y}(1+g)^{k-2025} = [(1+g)^{y-2024}-1]/g$\,PB,
so the annual electricity bill is:
\begin{equation}
  C_{\text{elec,HDD}}^{\text{full}}(y)
    = c_{\text{elec,HDD}}(y) \cdot S^{\text{full}}_{\text{data}}(y).
  \label{eq:celecfull}
\end{equation}

\smallskip
\noindent\textbf{Cumulative cost.}
The cumulative HDD electricity cost of materialized-artifact storage over $[2025,y]$ under
exponential growth is
\begin{align}
% steady (g=0) version commented out; only the growing (g) version is kept
%  E_{\text{full}}(y)
%    &= \sum_{\tau=2025}^{y} c_{\text{elec,HDD}}(\tau)\,(\tau-2024),
%  \label{eq:celecfullcum0} \\
  E_{\text{full}}^{(g)}(y)
    &= \sum_{\tau=2025}^{y} c_{\text{elec,HDD}}(\tau)\,S^{\text{full}}_{\text{data}}(\tau),
  \label{eq:celecfullcum}
\end{align}
with the linear case recovered as the $g\to0$ limit of the exponential form, and
$c_{\text{elec,HDD}}(\tau)$ now inside the sum since the unit cost declines year over year.
Adding this operational electricity to the replacement-inclusive capital
cost gives the total cost of materialized-artifact storage on HDD through year $y$:
\begin{equation}
  C_{\text{total}}^{(g)}(y)
    = C_{\text{HDD,cum}}^{(g)}(y) + E_{\text{full}}^{(g)}(y),
  \label{eq:chddtotal}
\end{equation}
with the linear baseline again its $g{=}0$ case,
$C_{\text{total}}\equiv C_{\text{total}}^{(0)}$.
The dollar totals at the 2045 horizon are reported in
\autoref{sec:horizon-cost}.

\subsection{Materialized-Artifact Storage on Tape}
\label{sec:tape-cost}

Magnetic tape is the natural cold-storage alternative to HDD, and it changes
the cost model in three ways.
First, tape media is rated for 15 to 30 years and is commonly quoted at
30\,years\cite{ltolifespan2024}, which exceeds our 2025--2045 horizon, so a
cartridge written at ingest survives the full study and is bought exactly once,
with no end-of-life replacement series of the kind that governs
\autoref{eq:provision}.
Second, the read and write mechanism is not bundled into the media as it is for
HDD but lives in separate tape drives and a robotic library, so tape carries
fixed infrastructure cost that HDD has no analogue for.
Third, cartridges sit unmounted in the library and draw no power, so only the
mounted drives and the library controller consume electricity, far below the electricity unit cost of
HDD $c_{\text{elec,HDD}}$ of~\autoref{sec:power-cost}.

\subsubsection{Tape Price Anchor}
\label{sec:tape-anchor}

We anchor on the current generation, LTO-10, which launched in mid-2025 at
30\,TB native per cartridge for about \$300\cite{fujifilm2025lto10}, giving a
native price of about \$10/TB\cite{techradar2025lto10}, so
$C_{\text{tape}}(2025)\approx 10{,}000$\,\$/PB, roughly half the HDD anchor of
20{,}000\,\$/PB.
As~\autoref{fig:tape-price} shows, native tape \$/TB has fallen steadily
across LTO-6 through LTO-10, from about \$47/TB in 2013 to about \$10/TB in
2025, measured from dated launch-era retail listings for each
generation\cite{tapeandmedia_ltoprices}.
A log-linear fit to these launch-era prices gives an annual decline of about
$12\%$, close to the HDD rate, so to keep tape on the same basis as HDD we adopt
a rounded $\mathrm{drop}_{\text{tape}}=10\%$ matching the
$\mathrm{drop}_{\text{HDD}}=10\%$ of~\autoref{sec:steady-1pb}.
The tape price therefore declines geometrically from its 2025 anchor,
$C_{\text{tape}}(t) = C_{\text{tape}}(2025)\,(1-\mathrm{drop}_{\text{tape}})^{(t-2025)}$.

\subsubsection{Media Cost}
\label{sec:tape-capital}

With one purchase per petabyte and a price that declines from the 2025 anchor,
the cumulative tape \emph{media} cost for exponential growth is
\begin{align}
% steady (g=0) version commented out; only the growing (g) version is kept
%  C_{\text{tape,cum}}(y)
%    &= \sum_{k=2025}^{y} C_{\text{tape}}(k),
%  \label{eq:ctapecum} \\
  C_{\text{tape,cum}}^{(g)}(y)
    &= \sum_{\tau=2025}^{y} S_{\text{data}}(\tau)\,C_{\text{tape}}(\tau),
  \label{eq:ctapecumg}
\end{align}
with no replacement multiplier, where
$C_{\text{tape}}(k)=C_{\text{tape}}(2025)\,(1-\mathrm{drop}_{\text{tape}})^{k-2025}$.
Over 2025--2045 the media alone costs
$C_{\text{tape,cum}}(2045)\approx\$89{,}000$ for linear growth and
$C_{\text{tape,cum}}^{(g)}(2045)\approx\$504{,}000$ for exponential growth.

\subsubsection{Drives, Library, and Power}
\label{sec:tape-infra}

Media is only part of the bill.
Reading or writing tape requires LTO-10 drives priced at over
\$10{,}000\,each\cite{spectra2025lto10drive}, which wear and are re-bought over
the horizon, and we provision them to the annual ingest rate.
A single drive sustains the LTO-10 native rate of $400$\,MB/s, so at a $50\%$
write duty it absorbs about $6.3$\,PB of ingest per year, and we hold a minimum
of two drives for redundancy.
Each cartridge of live capacity also needs a slot in a robotic library, whose
frame and robotics we charge at about $\$100$ per
slot\cite{tapelibrary2024}, a durable cost provisioned once as the corpus
grows.
Finally, only mounted drives and the library controller draw power, so we apply the electricity unit cost for tape $c_{\text{elec,tape}}(t) = 0.1\,c_{\text{elec,HDD}}(t)$
derived in~\autoref{sec:param-anchors}, \autoref{eq:alphatape},
anchored at $31.5$\,\$/PB/yr, a fixed $90\%$ below the HDD
$c_{\text{elec,HDD}}(2025) = 315$\,\$/PB/yr because idle cartridges consume
nothing.
The full tape cost is the sum of media, drives, library, and power, and the
resulting breakdown at the 2045 horizon is reported in
\autoref{sec:horizon-cost}.

\subsection{Cost Over Next 20 years}
\label{sec:horizon-cost}

We now evaluate both media over the 2025--2045 horizon and report the dollar totals for the linear and exponential growth. While we report storing data for 20 years, the results are similar for 5-year or shorter storage duration except tape cost increases due to the library cost is not fully amortized.

\noindent\textbf{HDD.}
Instantiating the replacement-inclusive capital of \autoref{eq:chddcumrepl} and the electricity of \autoref{eq:chddtotal} with the anchors of \autoref{sec:param-anchors} gives the totals in \autoref{tab:horizon-hdd}.
For the linear 1\,PB-per-year baseline ($g{=}0$), the replacement-inclusive
HDD capital is $\approx\$344{,}000$ and the operational electricity
$\approx\$21{,}000$, for a total of $C_{\text{total}}(2045)\approx\$365{,}000$.
For exponential growth ($g{=}20\%$), the same accounting yields
$\approx\$1{,}515{,}300$ of HDD capital and $\approx\$81{,}000$ of
electricity, a total of $C_{\text{total}}^{(g)}(2045)\approx\$1{,}597{,}000$,
about $4.4\times$ the linear baseline because the compounding ingest
front-loads drive purchases at today's higher prices while the powered corpus
accumulates geometrically.

\smallskip
\noindent\textbf{Tape.}
\autoref{tab:horizon-tape} reports the tape breakdown at the same horizon.
For linear growth, the full tape cost is around $\$261{,}000$; HDD costs about
$1.4\times$ as much. This advantage is smaller than the $4.1\times$ difference
that media alone would suggest because the fixed drive and
library cost is spread over little data.
For exponential growth, tape remains cheaper: at around
$\$1{,}422{,}000$, compared with the \$1{,}597{,}000 full-HDD total,
HDD costs about $1.1\times$ as much.
The one component that scales adversely is the library, which at one slot per
cartridge grows with the full corpus to about \$750{,}000 by 2045 and becomes
the largest single tape cost at scale.
Tape therefore remains the cheaper cold store across both the linear and exponential
regimes once drives, library, and power are accounted for, with its
per-terabyte edge preserved rather than eroded once the price trend is included.

\begin{table*}[t]
  \centering
  \footnotesize
  \caption{Cumulative cost at the 2045 horizon by component, under linear ($g{=}0$) and exponential ($g{=}20\%$ annual ingest growth) corpus growth. For materialized-artifact storage on HDD, capital expenditure dominates the total cost, with replacement adding at least half of the initial hardware cost. In contrast, tape media is purchased once and amortized over the full 30-year cartridge lifetime, so media contributes only a small fraction of total tape cost. Instead, tape libraries and drives account for more than half of the overall cost.}
  \label{tab:horizon-cost}
  \begin{subtable}[b]{0.48\textwidth}
    \centering
    \setlength{\tabcolsep}{6pt}
    \renewcommand{\arraystretch}{1.1}
    \begin{tabular}{@{}lrr@{}}
      \toprule
      HDD component & Linear ($g{=}0$) & Exponential ($g{=}20\%$) \\
      \midrule
      Capital, w/o replace & \$178{,}116 & \$1{,}008{,}458 \\
      Capital, w/ replace  & \$343{,}908 & \$1{,}515{,}300 \\
      Electricity          &  \$20{,}839 &    \$81{,}366 \\
      \midrule
      Total                & \$364{,}747 & \$1{,}596{,}666 \\
      \bottomrule
    \end{tabular}
    \caption{HDD}
    \label{tab:horizon-hdd}
  \end{subtable}
  \hfill
  \begin{subtable}[b]{0.48\textwidth}
    \centering
    \setlength{\tabcolsep}{6pt}
    \renewcommand{\arraystretch}{1.1}
    \begin{tabular}{@{}lrr@{}}
      \toprule
      Tape component & Linear ($g{=}0$) & Exponential ($g{=}20\%$) \\
      \midrule
      Media         &  \$89{,}058 &   \$504{,}229 \\
      Drives        & \$100{,}000 &   \$160{,}000 \\
      Library       &  \$70{,}000 &   \$750{,}100 \\
      Electricity         &   \$2{,}084 &     \$8{,}137 \\
      \midrule
      Total         & \$261{,}142 & \$1{,}422{,}466 \\
      \bottomrule
    \end{tabular}
    \caption{Tape}
    \label{tab:horizon-tape}
  \end{subtable}
\vspace{-15pt}
\end{table*}

\section{Cost Model 2: Prompt-Based Regeneration}
\label{sec:regen-cost-model}

\autoref{sec:hdd-cost-model} modeled the cost of retaining a corpus in full.
This section prices the opposite regime, in which each artifact is
regenerated by its generator whenever it is requested, and asks under what
conditions that regime is the cheaper one.
We first consider
prompt-based regeneration, then we add a hot cache backed by regeneration.
Both policies include capital and electricity costs and reduce to a
single decision variable: the break-even generation cost at which
regenerating on demand stops being cheaper than the materialized-artifact storage baseline on HDD.

\smallskip
\noindent\textbf{Access model.}
We assume that each artifact has size $s_{\text{obj}}$\,PB and receives ten
requests per year on average. The annual request rate in year $y$ is therefore
ten times the cumulative artifact count.
% For the steady 1\,PB ingest, the cumulative artifact count $N_{\text{obj}}(y)$
% and request rate $R(y)$ through year $y$ are:
% \begin{equation}
%   N_{\text{obj}}(y) = \frac{y - 2024}{s_{\text{obj}}}
%   \qquad\text{and}\qquad
%   R(y) = \frac{10\,(y-2024)}{s_{\text{obj}}}. \label{eq:nobj}
% \end{equation}
The cumulative artifact count is the running ingest divided by object size, so
with the corpus growing at $g$ per year (\autoref{sec:growing-corpus}) the
request rate is:
\begin{equation}
  R^{(g)}(y) 
    = \frac{10}{s_{\text{obj}}}\sum_{\tau=2025}^{y} S_{\text{data}}(\tau)
    = \frac{10}{s_{\text{obj}}}\sum_{j=0}^{y-2025}(1+g)^{j},
  \label{eq:nobjg}
\end{equation}

with the linear case recovered at $g=0$. We assume an independent reference model (IRM)~\cite{fagin1977asymptotic,fricker2012versatile}, in which each request independently selects an object from a fixed popularity distribution.

\smallskip
\noindent\textbf{AI-generated image.}
We ground the model in text-to-image generation, the running example for the
rest of this section. A modern diffusion generator such as FLUX.1-dev
generates an image from its text prompt by starting from pure Gaussian noise
and iteratively denoising it over a fixed schedule of steps, each step a full
forward pass through the multi-billion-parameter backbone, until the denoised output can be decoded into the final image. Regenerating one $1024\times1024$ image this way
costs $\text{flop}_{\text{regen,prompt}} = 2{,}100$\,TFLOP and reproduces an image at
$s_{\text{obj}} = 3$\,MB. This is the prompt-based regeneration pipeline
that the two policies below invoke on every request in
\autoref{sec:regen-full} and on every cache miss in
\autoref{sec:caching}.

\noindent\textbf{Compute price.}
Regeneration prices compute at the GPU anchor introduced in
\autoref{sec:cost_trends}: we take
$c_{\text{GPU}}(2025) = 18$\,[\$/(TFLOP/s)]\cite{epochai2024mlhardware}
and an annual decline $\mathrm{drop}_{\text{GPU}} = 30\%$, so
$c_{\text{GPU}}(t) = 18\times 0.70^{(t-2025)}$\,[\$/(TFLOP/s)].

\subsection{Prompt-Based Regeneration}
\label{sec:regen-full}
The simplest regeneration policy does not store any generated artifacts. Instead, every
request is served by rerunning the full prompt-to-image pipeline on demand. This
policy avoids both HDD capacity and HDD electricity, but it pays the full
generation cost each time an artifact is requested.

% The first policy stores nothing and regenerates every requested artifact
% from its prompt on demand.

\smallskip
\noindent\textbf{Regeneration cost.}
We model the GPU capacity price, measured in dollars per TFLOP/s of provisioned throughput, as an exponentially declining quantity. Amortizing this capacity cost over a GPU's service lifetime and effective
utilization gives the serving price per TFLOP:
\begin{equation}
  c_{\text{GPU}}(y)
    = c_{\text{GPU}}(2025)\,(1-\mathrm{drop}_{\text{GPU}})^{(y-2025)}
    \;[\$/(\text{TFLOP/s})], \label{eq:pgpu}
\end{equation}
\begin{equation}
  c_{\text{TFLOP}}(y)
    = \frac{c_{\text{GPU}}(y)}{L \cdot u \cdot \Delta_{\text{yr}}}
    \;[\$/\text{TFLOP}], \label{eq:ptflop}
\end{equation}
with GPU lifetime $L = 3$\,years, utilization $u=0.40$, and $\Delta_{\text{yr}}$
the number of seconds in a year.
The annual prompt-based regeneration cost is the compute spend required to serve all
requests in year $y$: each request consumes
$\text{flop}_{\text{regen,prompt}}$ TFLOP, and that computation is charged at the
year-$y$ serving price $c_{\text{TFLOP}}(y)$. Under exponential corpus growth, this gives:
% steady (g=0) version commented out; only the growing (g) version is kept
% \begin{equation}
%   C_{\text{regen}}(y) = R(y) \cdot \text{flop}_{\text{regen,prompt}} \cdot c_{\text{TFLOP}}(y),
%   \label{eq:cregen}
% \end{equation}
% and for the growing corpus:
\begin{equation}
  C_{\text{regen}}^{(g)}(y)
    = R^{(g)}(y) \cdot \text{flop}_{\text{regen,prompt}} \cdot c_{\text{TFLOP}}(y).
  \label{eq:cregeng}
\end{equation}

\smallskip
\noindent\textbf{Electricity.}
Prompt-based regeneration stores no materialized artifacts, so it draws no HDD power. The only
operational electricity consumption is GPU compute for every request, priced at $c_{\text{elec,GPU}}(t)$\,[\$/TFLOP]. The price is anchored at
$c_{\text{elec,GPU}}(2025) = 9.78\times10^{-9}$\,\$/TFLOP in year 2025, and declines at
$\mathrm{drop}_{\text{GPU}} = 30\%$/yr:
\begin{equation}
  C_{\text{elec,regen}}(y) = R^{(g)}(y)\,\text{flop}_{\text{regen,prompt}}\,c_{\text{elec,GPU}}(y).
  \label{eq:celecregen}
\end{equation}

\smallskip
\noindent\textbf{Break-even.}
Prompt-based regeneration is cheaper than materialized-artifact storage on HDD only if its
cumulative compute and GPU-electricity cost stays below the materialized-artifact storage cost
$C^{(g)}_{\text{total}}(y)$ from~\autoref{eq:chddtotal}. Equivalently,
through year $y$ the policy wins when: 
\begin{equation}
  \text{flop}_{\text{regen,prompt}} \;<\;
  \frac{C_{\text{total}}^{(g)}(y)}
       {\displaystyle\sum_{\tau=2025}^{y}
         R^{(g)}(\tau)\bigl(c_{\text{TFLOP}}(\tau)+c_{\text{elec,GPU}}(\tau)\bigr)}.
  \label{eq:cthreshpow}
\end{equation}

\smallskip
\noindent\textbf{Cost over the 20 years.}
For the running 3\,MB image example over 2025--2045, materialized-artifact storage on
HDD with replacement and electricity costs about $\$365{,}000$ under linear
1\,PB-per-year ingest and about $\$1.60$\,M under $20\%$ annual ingest growth.
Setting the cumulative regeneration spend of~\autoref{eq:cthreshpow}
equal to these totals yields a break-even generation cost of only about
$20$\,TFLOP for linear growth and about $50$\,TFLOP for exponential growth.
Every practical generator lies far above this bound, so prompt-based
regeneration is vastly more expensive: at FLUX.1-dev's $\text{flop}_{\text{regen,prompt}}=2{,}100$\,TFLOP the
compute and GPU-electricity bill reaches about $\$38$\,M for linear growth
and about $\$68$\,M for exponential growth, far above the cost of
materialized-artifact storage.

\begin{takeaway}
\textbf{Takeaway 1:} Materialized-artifact storage over 20 years costs
about $100\times$ less than prompt-based regeneration for linear growth and about
$40\times$ less for exponential growth, since a stored artifact is written once while
regeneration repays its full cost on every request. % \jason{talk about 5 years}
\end{takeaway}

\subsection{Caching Hot Objects}
\label{sec:caching}
Regeneration without storage is expensive because every request incurs the full prompt-to-artifact cost. Accesses, however, are not uniform: a small set of popular objects often receives most requests. A hot-object cache avoids repeated regeneration for these objects and confines regeneration to cache misses. We therefore extend the cost model to include a cache.

Let $h$ be the target hit ratio, the fraction of requests served directly
from the cache, and let $F_{\text{cached}}$ be the fraction of the corpus
that must be cached to reach that hit ratio.
The mapping from $h$ to $F_{\text{cached}}$ is determined by the underlying popularity distribution. We use a Zipfian workload, a request-popularity model repeatedly observed in production caching traces~\cite{breslau1999zipf,yang2020twitter}. As shown in \autoref{app:access-pattern}, a hit ratio of $h = 0.90$ requires a cached fraction of $F_{\text{cached}} \approx 0.944\,N^{-0.10}$. Thus, achieving a 90\% hit ratio requires caching about $15\%$ of the artifacts in a corpus of $N = 10^{8}$ and about $9.4\%$ in a corpus of $N = 10^{10}$.

%  we instantiate it for a Zipfian workload in
% ~\autoref{sec:recompute} and treat both $h$ and $F_{\text{cached}}$ as
% given inputs here.

\smallskip
\noindent\textbf{Cache size and volume.}
With the cumulative object count $N(y)$ from~\autoref{eq:nobjg}, the
cache holds $F_{\text{cached}}\,N(y)$ artifacts, so multiplying by the object
size $s_{\text{obj}}$ gives its physical volume in PB:
\begin{equation}
  S_{\text{cache,data}}(y) = F_{\text{cached}}\,N(y)\,s_{\text{obj}}.
  \label{eq:vcache}
\end{equation}
A skewed workload keeps $F_{\text{cached}}$ small, so a modest cache absorbs
the majority of requests even as the catalog grows large.

\smallskip
\noindent\textbf{Annual HDD cost for the growing cache.}
As the corpus grows, the cache also expands. In year $y$, the operator
therefore buys only the additional cache capacity needed since the
previous year, defined as $\Delta S_{\text{cache,data}}(y) = S_{\text{cache,data}}(y) - S_{\text{cache,data}}(y-1)$, together with replacement capacity for cache drives that were
installed $L_{\text{HDD}}$ years earlier. Combining these two capacity requirements gives a total annual purchase of
$S_{\text{cache,purchase}}(y)=\Delta S_{\text{cache,data}}(y)+S_{\text{cache,purchase}}(y-L_{\text{HDD}})$ and the following HDD cost for the cache:
\begin{equation}
  C_{\text{cache,HDD}}(y)
    = S_{\text{cache,purchase}}(y)\cdot C_{\text{HDD}}(y).
  \label{eq:ccachehdd}
\end{equation}

\smallskip
\noindent\textbf{Regeneration cost for cache misses.}
The cache serves fraction $h$ of requests, leaving fraction $1-h$ to be
regenerated. Each miss consumes $\text{flop}_{\text{regen,prompt}}$ TFLOP and
is charged at the year-$y$ serving price $c_{\text{TFLOP}}(y)$, giving
\begin{equation}
  C_{\text{miss}}(y)
    = (1-h)\,R^{(g)}(y)\cdot \text{flop}_{\text{regen,prompt}}\cdot c_{\text{TFLOP}}(y).
  \label{eq:cmiss}
\end{equation}
Before adding electricity, the annual cost of the cache policy is therefore:
\begin{equation}
  C_{\text{mod}}(y) = C_{\text{cache,HDD}}(y) + C_{\text{miss}}(y).
  \label{eq:cmod}
\end{equation}

\smallskip
\noindent\textbf{Electricity.}
The cache consumes HDD electricity, while cache-miss regeneration consumes GPU electricity. Their respective costs are
\begin{equation}
  C_{\text{elec,cache}}(y) = c_{\text{elec,HDD}}(y)\,S_{\text{cache,data}}(y),
  \qquad
  C_{\text{elec,GPU}}(y) = (1-h)\,R^{(g)}(y)\,\text{flop}_{\text{regen,prompt}}\,c_{\text{elec,GPU}}(y).
  \label{eq:celecgpu}
\end{equation}

\smallskip
\noindent\textbf{Break-even.}
Counting capital and electricity on both sides, caching beats full HDD
storage when the cache cost plus miss regeneration, including powered-cache
and miss-compute electricity, stays below the total storage cost
$C_{\text{total}}^{(g)}(y)$. The new break-even threshold $\text{flop}^{*}_{\text{regen,prompt}}$ is updated to:
\begin{equation}
  \text{flop}_{\text{regen,prompt}} \;<\;
  \frac{C_{\text{total}}^{(g)}(y)
        - \displaystyle\sum_{\tau=2025}^{y}
          \bigl[C_{\text{cache,HDD}}(\tau)+C_{\text{elec,cache}}(\tau)\bigr]}
       {\displaystyle\sum_{\tau=2025}^{y}
         (1-h)\,R^{(g)}(\tau)\bigl(c_{\text{TFLOP}}(\tau)+c_{\text{elec,GPU}}(\tau)\bigr)}.
  \label{eq:cachethreshpow}
\end{equation}
Artifacts whose generation cost $\text{flop}_{\text{regen,prompt}}$ falls \emph{below} this
break-even are cheap enough to regenerate on miss.

\subsection{Cost Over Next 20 years}
A Zipfian hot cache at $h=0.90$ holds only a small fraction of the corpus,
costing about $\$38{,}000$ for linear growth and about $\$137{,}000$ for
exponential growth in cache HDD plus its electricity.
Measured against the same $\$365{,}000$ and $\$1.60$\,M materialized-artifact storage baselines,
\autoref{eq:cachethreshpow} raises the break-even to about $182$\,TFLOP
for linear growth and about $454$\,TFLOP for exponential growth.
Confining regeneration to the $10\%$ of requests that miss lifts the threshold
roughly tenfold over prompt-based regeneration, yet FLUX.1-dev at $2{,}100$\,TFLOP still
sits well above it, so its misses remain too expensive to regenerate at the full size.

\begin{takeaway}
\textbf{Takeaway 2:} A hot cache that regenerates only the cold misses reduces cost, however, it does not beat materialized-artifact storage. 
\end{takeaway}

\section{Cost Model 3: LazIR, an Intermediate-Representation Store}
\label{sec:cache-ir}

\begin{figure}[t]
  \centering
  \includegraphics[width=0.8\columnwidth]{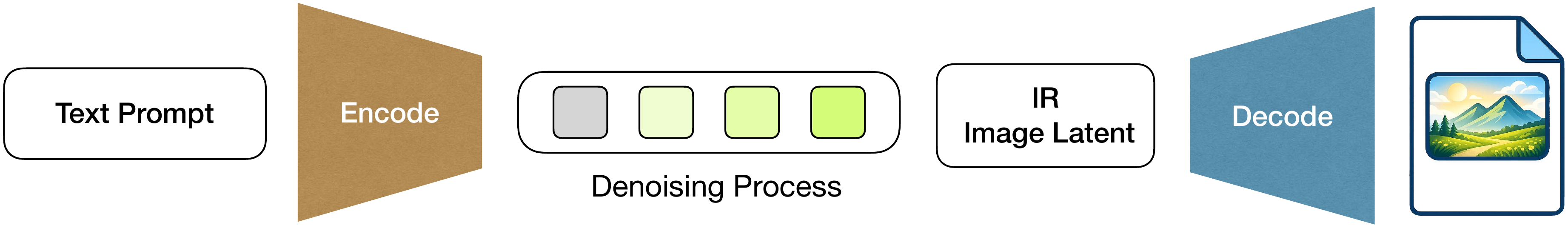}
  \vspace{-5pt}
  \caption{Intermediate representation for text-to-image generation. Instead of
           persisting the decoded pixels, LazIR stores the compact
           \emph{IR} (about $1/6$ the size for FLUX.1-dev, with
           other models requiring only $1/24$ or $1/48$) and
           regenerates the image on demand with a single VAE decode, skipping
           the expensive iterative denoising. Transforming from the IR is far
           cheaper than prompt-based regeneration: the decode costs about
           $10.3$\,TFLOPs versus $2{,}100$\,TFLOPs for prompt-based generation, a $1/200$
           compute fraction.}
  \vspace{-10pt}
  \label{fig:ir-image}
\end{figure}

\noindent\textbf{Diffusion models.}
Prompt-based regeneration is not the only choice for regenerating the artifact. Here we focus on image generation which relies on diffusion models and generate in three steps: first, it encodes the prompt and / or input image into a compact \textit{latent}, then performs multiple steps (e.g., 48) of denoising, and last it decodes the denoised latent back to pixels. The encode and decode steps are often not compute intensive, while the denoise process accounts of majority of the computation. 
Persisting the compact latent as an \emph{intermediate representation(IR)} allows us to save space and skip the expensive denoising (\autoref{fig:ir-image}).

% Because the decoder reproduces the image from
% the same latent the model itself would have decoded, fidelity is limited only
% by the VAE's reconstruction quality and is effectively indistinguishable from
% the originally generated image.

\smallskip
\noindent\textbf{Regeneration from IR.}
We now refine the storage model of \autoref{sec:caching} by replacing prompt-based regeneration with \emph{LazIR}: every object ever ingested is
persisted as a compact intermediate representation of size $k_{\text{size}} s_{\text{obj}}$
($0 < k_{\text{size}} \le 1$), while the hot working set is also cached. Only cache misses are regenerated from the stored latent at reduced compute cost $k_{\text{gen}} \text{flop}_{\text{regen,prompt}}$ per access ($0 < k_{\text{gen}} \le 1$).
For an FP16 latent with $c$ channels and spatial downsampling factor $f$, relative
to a decoded 8-bit RGB image of the same resolution, the store fraction is
$k_{\text{size}}=2c/(3f^2)$. FLUX.1-dev uses $f{=}8$ and $c{=}16$, giving
$k_{\text{size}}=1/6$ (a $6\times$ reduction), the conservative image setting
used in our evaluation~\cite{labs2025flux1kontextflowmatching}. Other widely used
models have substantially smaller latents: Stable Diffusion 1.5/2.x and SDXL use
$f{=}8,c{=}4$, giving $k_{\text{size}}=1/24\approx0.042$ (a $24\times$
reduction)~\cite{rombach2022high,podell2023sdxl}, while SANA's DC-AE uses
$f{=}32,c{=}32$, giving $k_{\text{size}}=1/48\approx0.021$ (a $48\times$
reduction)~\cite{xie2025sana}. Regenerating the image no longer
re-runs the denoising schedule, it is a single forward pass through the VAE
decoder costing about $10.3$\,TFLOP, a decode fraction $k_{\text{gen}} = 1/200$ of the
$2{,}100$\,TFLOPs full generation.

\smallskip
The modified annual cost now comprises three components:
\begin{equation}
  C_{\text{mod}}^{(\text{IR})}(y)
    = C_{\text{cache,HDD}}(y)
    + C_{\text{IR,HDD}}(y)
    + C_{\text{miss,IR}}(y),
  \label{eq:cmodir}
\end{equation}
where:
\begin{itemize}[noitemsep,leftmargin=*]
  \item $C_{\text{cache,HDD}}(y)$ is the HDD cost for the hot cache
    (\autoref{eq:ccachehdd}) for artifacts in materialized form.
  \item $C_{\text{IR,HDD}}(y) = \,k_{\text{size}} S_{\text{HDD,purchase}}(y)
    \cdot C_{\text{HDD}}(y)$ is the annual HDD spend on the growing
    latent store covering \emph{all} artifacts, where
    \begin{equation}
      S_{\text{IR,data}}(y) = N(y)\cdot k_{\text{size}} s_{\text{obj}}
      \label{eq:vir}
    \end{equation}
    is the total latent-store volume in PB.
    The store grows by $k_{\text{size}}(1{+}g)^{y-2025}$\,PB of new artifacts each
    year and is replaced on the same $L_{\text{HDD}}$ cycle, so its
    annual purchase is exactly a fraction $k_{\text{size}}$ of the full-corpus
    provisioning $S_{\text{HDD,purchase}}(y)$.
  \item $C_{\text{miss,IR}}(y)
    = (1-h)\,R^{(g)}(y)\cdot k_{\text{gen}} \text{flop}_{\text{regen,prompt}}\cdot c_{\text{TFLOP}}(y)$
    is the regeneration cost for cache misses, now reduced by
    a factor of $k_{\text{gen}}$ because each miss decodes from the stored
    intermediate representation rather than running the full
    denoising regeneration.
\end{itemize}
With the IR store provisioned as $k_{\text{size}} S_{\text{HDD,purchase}}(\tau)$ each year and priced
at $C_{\text{HDD}}(\tau)$, the cumulative latent-store HDD bill is exactly
a fraction $k_{\text{size}}$ of the materialized-artifact storage baseline, replacements
included:
\begin{align}
  \sum_{\tau=2025}^{y} C_{\text{IR,HDD}}(\tau)
    &= k_{\text{size}}\cdot C_{\text{HDD,cum}}^{(g)}(y).
  \label{eq:cirhddsumapprox}
\end{align}

\smallskip
\noindent\textbf{Electricity.}
LazIR has both storage and compute electricity terms. The powered HDD volume consists of the IR store and the full-fidelity hot cache, for a total capacity of $k_{\text{size}} S^{\text{full}}_{\text{data}}(y) + S_{\text{cache,data}}(y)$. Cache misses are regenerated at the reduced cost $k_{\text{gen}} \text{flop}_{\text{regen,prompt}}$:
\begin{equation}
  \begin{aligned}
    C_{\text{elec,HDD}}^{\text{LazIR}}(y)
      &= c_{\text{elec,HDD}}(y)\bigl[k_{\text{size}} S^{\text{full}}_{\text{data}}(y) + S_{\text{cache,data}}(y)\bigr], \\
    C_{\text{elec,GPU}}^{\text{IR}}(y)
      &= (1-h)\,R^{(g)}(y)\,k_{\text{gen}} \text{flop}_{\text{regen,prompt}}\,c_{\text{elec,GPU}}(y).
  \end{aligned}
  \label{eq:celec2tier}
\end{equation}
The cumulative  electricity cost for HDD and GPU computation over $[2025,y]$ is:
\begin{equation}
  \begin{aligned}
    E_{\text{LazIR}}(y)
      ={}& k_{\text{size}}\,E_{\text{full}}^{(g)}(y)
        + \sum_{\tau=2025}^{y} C_{\text{elec,cache}}(\tau) \\
      &+ k_{\text{gen}} \text{flop}_{\text{regen,prompt}} \sum_{\tau=2025}^{y} (1-h)\,R^{(g)}(\tau)\,c_{\text{elec,GPU}}(\tau).
  \end{aligned}
  \label{eq:celec2tiercum}
\end{equation}

\smallskip
\noindent\textbf{Break-even.}
Counting capital and electricity, LazIR is cheaper than materialized-artifact storage when the savings from shrinking the IR store exceed the costs of the hot cache and miss regeneration. Solving this condition for the full-generation cost gives
% substituting
% \autoref{eq:cirhddsumapprox} and \autoref{eq:cmiss} (with $k_{\text{gen}} \text{flop}_{\text{regen,prompt}}$ in
% place of $\text{flop}_{\text{regen,prompt}}$) and the two-tier electricity $E_{\text{LazIR}}(y)$
% into $\sum_\tau C_{\text{mod}}^{(\text{IR})}(\tau)
% + E_{\text{LazIR}}(y) < C_{\text{total}}^{(g)}(y)$ and
% rearranging, and using that both GPU capital ($c_{\text{TFLOP}}$) and GPU
% electricity ($c_{\text{elec,GPU}}(\tau)$) scale with $k_{\text{gen}} \text{flop}_{\text{regen,prompt}}$, 
\begin{equation}
  \text{flop}_{\text{regen,prompt}} \;<\;
  \frac{(1-k_{\text{size}})\,C_{\text{total}}^{(g)}(y)
        -\displaystyle\sum_{\tau=2025}^{y}
          \bigl[C_{\text{cache,HDD}}(\tau)+C_{\text{elec,cache}}(\tau)\bigr]}
       {k_{\text{gen}}\displaystyle\sum_{\tau=2025}^{y}
         (1-h)\,R^{(g)}(\tau)\bigl(c_{\text{TFLOP}}(\tau)+c_{\text{elec,GPU}}(\tau)\bigr)}.
  \label{eq:cachethreshirpow}
\end{equation}
The numerator is reduced by a factor $(1-k_{\text{size}})$ for storing the IR and
the denominator shrinks by $k_{\text{gen}}$, lifting the threshold by approximately
$(1-k_{\text{size}})/k_{\text{gen}}$ relative to the plain cache of
\autoref{eq:cachethreshpow}.

\smallskip
\noindent\textbf{Cost over the 20 years.}
% Storing a compact intermediate representation for every object and decoding
% each miss at the reduced cost $k_{\text{gen}} \text{flop}_{\text{regen,prompt}}$ lifts the threshold by about
% $\tfrac{1-k_{\text{size}}}{k_{\text{gen}}}$.
For FLUX.1-dev with $k_{\text{size}}=1/6$ and $k_{\text{gen}}=1/200$,
\autoref{eq:cachethreshirpow} gives a break-even of about
$29{,}700$\,TFLOPs for linear growth and about $74{,}200$\,TFLOPs for
exponential data growth. 
FLUX's actual $2{,}100$\,TFLOPs now sits well over an order of magnitude below the
threshold, so LazIR serves the entire workload for about
$\$118{,}000$ under linear growth and about $\$437{,}000$ under exponential growth, against $\$365{,}000$ and
$\$1.60$\,M (3.1$\times$-3.7$\times$) for materialized-artifact storage.

\begin{takeaway}
\textbf{Takeaway 3:} Storing compact intermediate representation and
decoding on miss lifts the image break-even to about $74{,}200$\,TFLOP, above
every current model. Together with caching, it reduces cost by $3.1\times$ to $3.7\times$ compared to materialized-artifact storage on HDD for linear and exponential growth.
\end{takeaway}

\section{Evaluation: Storage versus Regeneration}
\label{sec:recompute}

We now instantiate the cost model for two generative modalities with substantially different generation costs and IR characteristics: image and music generation. The materialized-artifact storage baseline is fixed by the corpus and is therefore identical across
modalities. Differences in the break-even points follow from each modality's object size $s_{\text{obj}}$,
per-object generation cost $\text{flop}_{\text{regen,prompt}}$, latent-store fraction $k_{\text{size}}$, and decode
fraction $k_{\text{gen}}$.
\autoref{tab:modality-ex} fixes one representative model per modality,
FLUX.1-dev~\cite{labs2025flux1kontextflowmatching} for images and
MusicGen-Large~\cite{copet2023musicgen} for music,
and summarizes the per-modality setup. We derive the music-specific values
below.

\begin{table}[t]
  \centering
  \small
  \setlength{\tabcolsep}{4pt}
  \caption{Per-modality example parameters.
           $s_{\text{obj}}$: object size; $\text{flop}_{\text{regen,prompt}}$: per-object generation cost;
           $k_{\text{size}}$: reduction ratio of intermediate representation size; $k_{\text{gen}}$: reduction ratio between IR-based regeneration and prompt-based regeneration.}
  \vspace{-5pt}
  \label{tab:modality-ex}
  \begin{tabular}{@{}llrrrrrr@{}}
    \toprule
    Modality & Model
    & $s_{\text{obj}}$
    & $\text{flop}_{\text{regen,prompt}}$ 
    & $k_{\text{size}}$
    & $k_{\text{gen}}$
    & \shortstack{Decode time\\(from IR)}
    & \shortstack{Prompt-based\\regeneration time} \\
    \midrule
    Image    & FLUX.1-dev     & 3\,MB   & 2{,}100 TFLOP & 1/6   & 1/200 & 40 ms  & 8{,}000 ms \\
    Music    & MusicGen-Large & 5.1\,MB & 21 TFLOP      & 1/212 & 1/470 & 0.4 ms & 188 ms \\
    \bottomrule
  \end{tabular}
  \vspace{-5pt}
\end{table}

\begin{figure}[t]
  \centering
  \includegraphics[width=0.8\columnwidth]{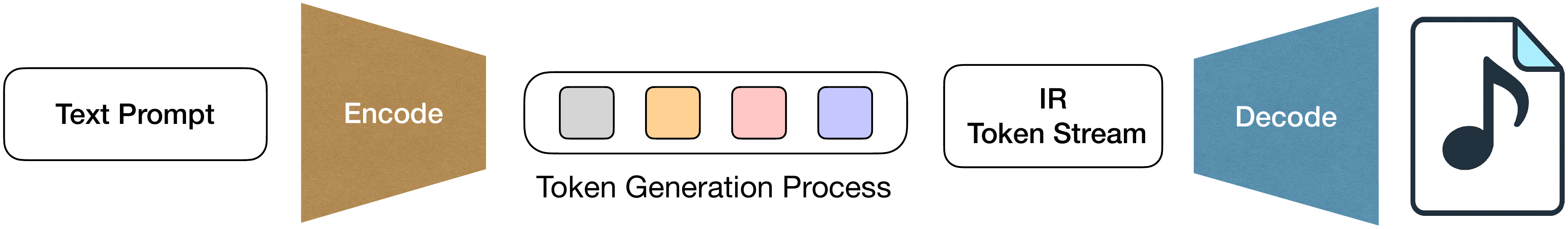}
  \vspace{-5pt}
  \caption{Intermediate representation for text-to-music generation.
           MusicGen-Large emits an EnCodec \emph{token stream} that LazIR stores
           as the IR (about $1/212$ the size of the decoded clip); on a miss a
           single non-autoregressive EnCodec decode regenerates the waveform.
           Regeneration from the IR costs about $0.045$\,TFLOP versus
           $21$\,TFLOP for autoregressive prompt-based regeneration, a $1/470$
           compute fraction.}
  \vspace{-5pt}
  \label{fig:ir-music}
\end{figure}

\noindent\textbf{Music with MusicGen-Large.}
MusicGen-Large does not predict the raw waveform sample by sample. Instead, a
neural audio codec called EnCodec~\cite{defossez2022encodec} first compresses sound into a short sequence
of discrete audio tokens, each token drawn from a learned vocabulary of sound
fragments, and the model then generates music by predicting these tokens much as
a language model predicts words.
% It is a 3.3B-parameter autoregressive transformer, so it emits the clip one
% frame at a time, every frame carrying 8 tokens from 8 residual codebooks that
% successively refine the sound, and a delay-interleaved pattern lets it predict
% all 8 codebooks of a frame together, taking about $1{,}507$ forward passes for
% the clip.
To keep the music faithful to the text prompt it applies classifier-free
guidance, forming both a prompted and an unprompted prediction at each step; the resulting generation costs around $21$\,TFLOP per clip.
% , of
% which $19.9$\,TFLOPs is weight multiply-adds, $0.9$\,TFLOPs the growing self-attention KV
% cache that carries earlier frames forward, and $0.6$\,TFLOPs the cross-attention that
% reads the prompt.
For example, a 30-second clip stored as uncompressed 44.1\,kHz stereo 16-bit PCM occupies
$S{=}5.1$\,MB. For model-generated music, the compact representation worth storing is the EnCodec token stream itself: eight residual
vector-quantization codebooks at 50 frames per second, so the whole clip is only
$30\times50\times8{=}12{,}000$ \texttt{int16} indices totaling $0.024$\,MB, a
store fraction $k_{\text{size}}{\approx}1/212$ (\autoref{fig:ir-music}).
Regenerating on a miss then costs only a small amount: the tokens are passed to the
convolutional EnCodec decoder, which regenerates the waveform in a single
non-autoregressive pass of about $0.045$\,TFLOP, giving $k_{\text{gen}}{\approx}1/470$.

\begin{figure}[t]
  \centering
  \includegraphics[width=0.8\columnwidth]{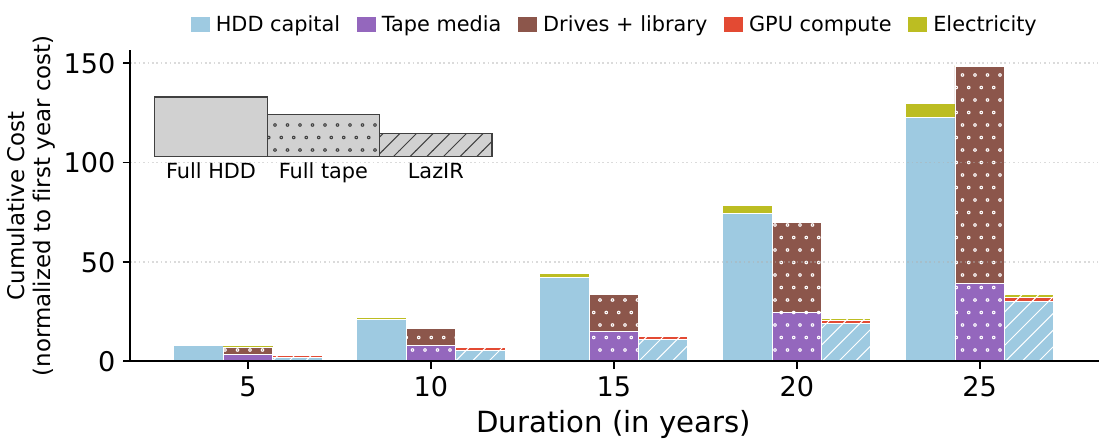}
  \label{fig:stack-image}
  % (b) total-cost-per-year line chart temporarily disabled
  % \begin{subfigure}{0.49\columnwidth}
  %   \includegraphics[width=\textwidth]{figures/cumcost_lines.pdf}
  %   \caption{Total cost per year, all modalities.}
  %   \label{fig:cost-lines}
  % \end{subfigure}
  \vspace{-10pt}
  \caption{Cumulative cost breakdown for serving FLUX.1-dev image corpora under 20\% annual ingest growth. All costs are normalized to the first-year cost of the materialized-artifact storage baseline and accumulated from 2025 onward. Tape is cheaper than HDD for retention periods shorter than 20 years, but its drive and library costs cause it to exceed HDD cost at 25 years. LazIR consistently achieves the lowest cost across all retention periods.}
           % The HDD and two-tier bars stack into HDD
           % capital, amortised GPU compute, and electricity, while the tape bar
           % splits into cartridge media, operational infrastructure of drives
           % plus library, and a near-zero electricity slice since idle cartridges
           % draw no power. Because the per-unit electricity of HDD storage
           % declines about 10\% per year and that of GPU compute declines about
           % 30\% per year, both exponentially, the compute and electricity slices
           % shrink over the horizon, so the later bars rise far more slowly than
           % the corpus itself.}
  \label{fig:modality-stack}
  \vspace{-15pt}
\end{figure}

\subsection{Cost composition over time}
\label{sec:cost-composition}
\autoref{fig:modality-stack} decomposes the cumulative cost at $g{=}20\%$,
contrasting the materialized-artifact storage baselines against the
LazIR design.
% The cumulative cost for image regeneration is broken into components including the capital, power and maintenance cost. 
Materialized-artifact storage on HDD costs about
$75\times$ the first-year materialized-artifact storage cost in capital and $4\times$ in
electricity by 2045.
Tape is a cheaper baseline, reaching about $70\times$ the
first-year cost by 2045, and its composition differs from HDD's: the
tape bar splits into roughly $25\times$ in cartridge media, a larger operational
slice of drives plus robotic library near $45\times$ within which the library
alone grows to about $37\times$ and becomes the dominant tape cost at scale, and
an almost invisible electricity slice since idle cartridges draw no power.
The tape total remains below the HDD baseline but well above the cost of LazIR. Both materialized-artifact storage baselines are identical across modalities because the corpus volume is fixed.

Each LazIR bar consists of a cache, a compact IR store, and the GPU cost of regenerating cache misses. The hot-cache cost is fixed by the corpus size and access skew, while the IR-store and regeneration costs shrink according to the store fraction $k_{\text{size}}$ and decode fraction $k_{\text{gen}}$. Thus, LazIR replaces most materialized-artifact storage with a much smaller IR store, and miss regeneration remains a small component in \autoref{fig:modality-stack} because decoding from an IR is inexpensive. The same pattern holds across modalities, with different $k_{\text{size}}$ and $k_{\text{gen}}$ values changing only the height of the LazIR bar: LazIR costs settle
near $27\%$ of the HDD baseline for images and near $10\%$ for music. The music result is omitted from the figure because it follows a trend similar to the image result.
% , the gap widening as the latent fraction $k_{\text{size}}$ shrinks.

\subsection{Regeneration cost across policies}
\label{sec:regen-cost}
\begin{figure}[t]
  \centering
  \includegraphics[width=0.8\columnwidth]{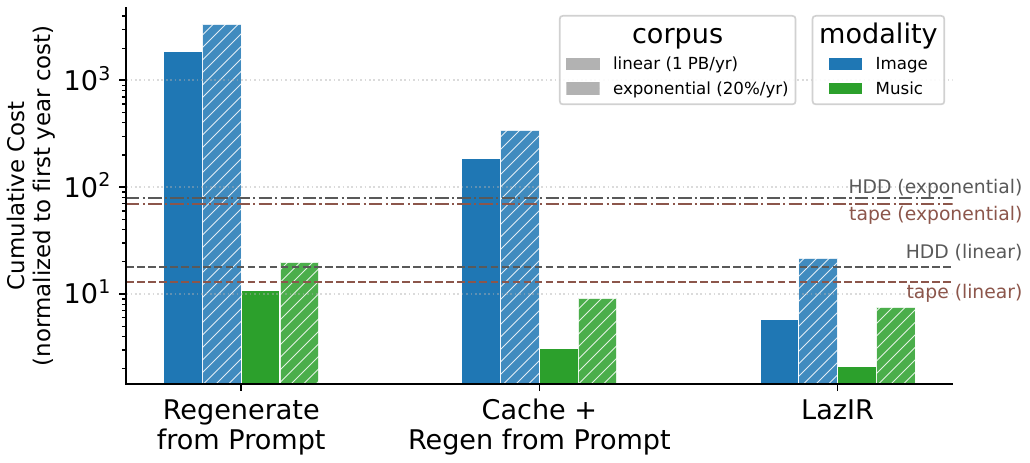}
  \vspace{-5pt}
  \caption{Cumulative cost from 2025 to 2045
           for the three regeneration policies applied to each modality under linear and exponential corpus growth,
           with every value normalized to the first-year materialized-artifact storage cost. The horizontal lines mark the materialized-artifact storage baselines, gray for HDD and brown for tape. Serving from stored IR reduces cost the most, consistently outperforming materialized-artifact storage across both modalities.}
  \vspace{-10pt}
  \label{fig:policy-cost}
\end{figure}

\autoref{fig:policy-cost} reports the cumulative 2025--2045 serving cost of
each modality under the three regeneration policies, normalized to the first-year
materialized-artifact storage cost, with the dashed lines marking the materialized-artifact storage
baselines.
Each storage medium contributes two baselines, one per growth regime: the gray HDD
lines sit at $18\times$ the first-year cost for linear growth and $79\times$
for exponential growth, while the brown tape lines sit lower, at about $13\times$ when data grows linearly and $70\times$ for exponential growth.
% , since tape media runs at roughly half the HDD
% price per terabyte and idle cartridges draw almost no power.
% Tape is therefore the tougher target a regeneration policy must beat, and every
% comparison below is read against both media.
Prompt-based regeneration is prohibitively expensive for images, exceeding both baselines by
one to two orders of magnitude, because the small per-object size yields an
enormous request count and every request pays the full generation cost.
Only music, which is cheap to regenerate, falls near the HDD line, though it
still sits above the lower tape line.
A hot cache narrows the gap by regenerating only the misses, yet its regeneration cost exceeds storage for images,
while its cost for music drops below the HDD line.
Only the LazIR design places every modality well below both baselines,
confirming it as the single policy that wins across both modalities and
both growth regimes, given a $8.6\times$ and $10.5\times$ reduction over linear and exponential growth respectively.

\subsection{Sensitivity of the break-even}
\label{sec:sensitivity}
\begin{figure}[t]
  \centering
  \includegraphics[width=0.45\columnwidth]{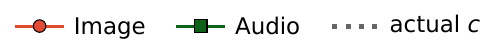}\\[2pt]
  \begin{subfigure}{0.32\columnwidth}
    \includegraphics[width=\linewidth]{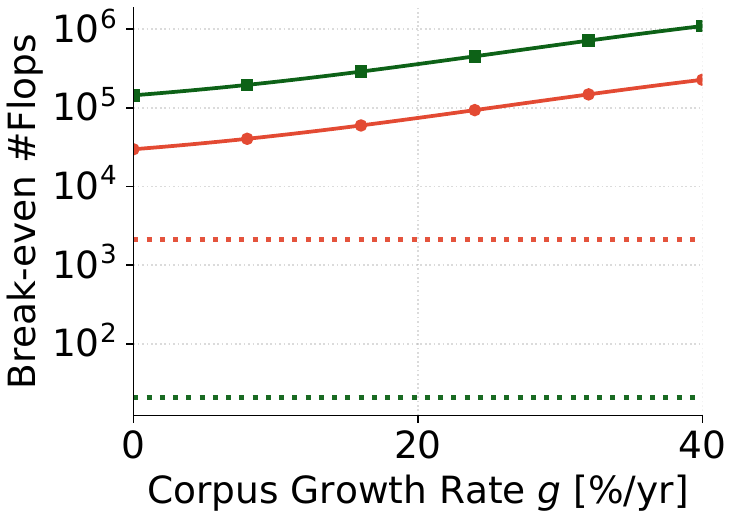}
    \caption{vs.\ ingest growth $g$}
    \label{fig:abl-g}
  \end{subfigure}%
  \hspace{2pt}%
  \begin{subfigure}{0.32\columnwidth}
    \includegraphics[width=\linewidth]{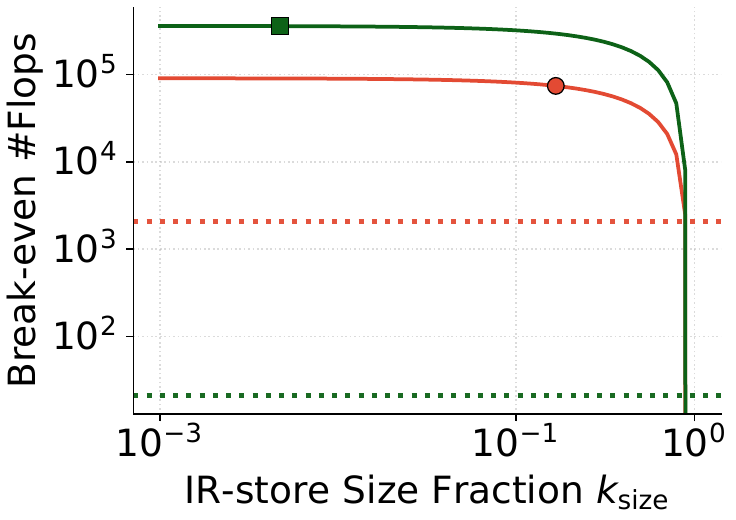}
    \caption{vs.\ store $k_{\text{size}}$}
    \label{fig:abl-k1}
  \end{subfigure}%
  \hspace{2pt}%
  \begin{subfigure}{0.32\columnwidth}
    \includegraphics[width=\linewidth]{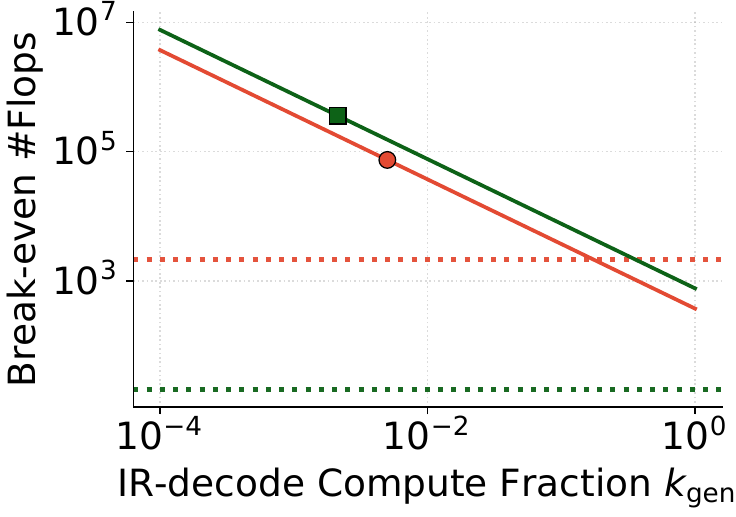}
    \caption{vs.\ decode $k_{\text{gen}}$}
    \label{fig:abl-k2}
  \end{subfigure}
  \vspace{-5pt}
  \caption{LazIR break-even FLOPs under varying annual growth rate, IR size, and regeneration-cost reduction. LazIR is better when the regeneration FLOPS is lower than the curve. We sweep one parameter at a time while holding the others fixed at each modality's operating point. Solid curves show the break-even FLOPs for prompt-based regeneration. Filled markers indicate the operating points for $k_{\text{size}}$ and $k_{\text{gen}}$, while dotted lines show each modality's actual generation cost. }
  \label{fig:ablation}
\end{figure}

\autoref{fig:ablation} sweeps the three drivers of the break-even
$\text{flop}^{*}_{\text{regen,prompt}}$, the generation cost below which IR-based regeneration beats materialized-artifact storage.
% holding the others at each modality's operating point.
The threshold rises with the annual ingest growth rate, by roughly $130\%$
as $g$ climbs from $0$ to $40\%$, because faster growth front-loads HDD
purchases at today's higher prices while the request load that drives
regeneration accumulates more gradually.
Against the store fraction $k_{\text{size}}$, the threshold is flat while the
intermediate representation is small and then collapses as $k_{\text{size}}$ approaches
one, near $k_{\text{size}}{\approx}0.9$, where the cold store costs almost as much as
keeping the materialized artifact and no finite generation cost can compensate.
For the decode fraction $k_{\text{gen}}$, the threshold scales inversely, so a tenfold
cheaper decode lifts the break-even tenfold, since the only $\text{flop}_{\text{regen,prompt}}$-dependent term
is the miss regeneration and it is proportional to $k_{\text{gen}} \text{flop}_{\text{regen,prompt}}$.
In both cases, the break-even thresholds at the operating points of both modalities sit far above their
actual generation costs, by one to several orders of magnitude, so LazIR
design retains a wide margin throughout the studied range.
% \yunjia{more explanation about the curve's shape?}

\section{Analysis on a Real-World Image Generation Workload}
\label{sec:ir-replay}

\begin{figure}[t]
  \centering
  \begin{subfigure}{0.32\columnwidth}
    \includegraphics[width=\textwidth]{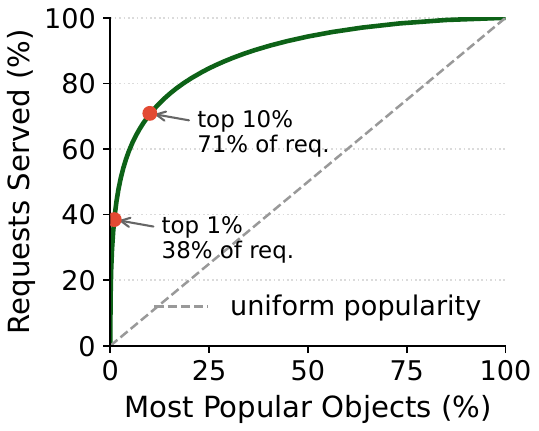}
    \caption{Request concentration.}
    \label{fig:civitai-pop}
  \end{subfigure}
  \hspace{2em}
  % \hfill
  \begin{subfigure}{0.48\columnwidth}
    \includegraphics[width=\textwidth]{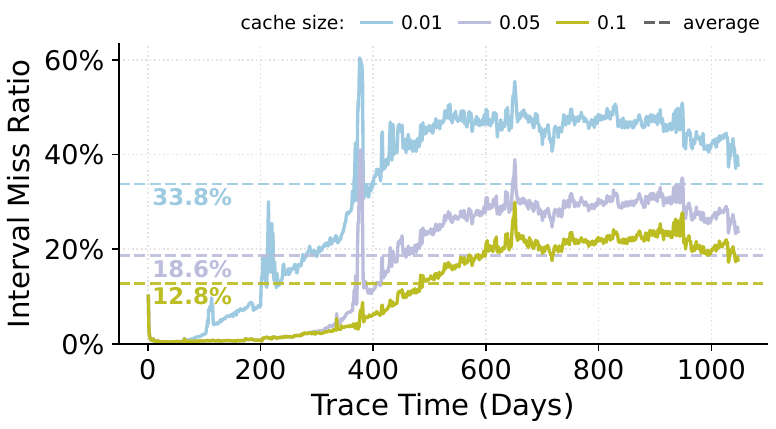}
    \caption{LRU per-interval miss ratio.}
    \label{fig:civitai-miss}
  \end{subfigure}
  \vspace{-5pt}
  \caption{Real-world generated-image access traces are highly skewed and cache-friendly. A small hot set serves most requests: the top $1\%$ and $10\%$ of images account for $38\%$ and $71\%$ of requests, respectively. An LRU cache further exploits temporal reuse on top of this popularity skew. With a cache sized to $10\%$ of distinct images, the whole-trace miss ratio drops to $12.8\%$, indicated by the dashed lines.}
  \label{fig:civitai-trace}
  \vspace{-10pt}
\end{figure}

We ground the previous analysis in a production access trace collected from CompanyX, a
large image generation and sharing platform~\footnote{We will open-source our trace after paper publication}.

\subsection{Trace Analysis}
The trace spans $2.87$ years, from April 2023 to March 2026, and contains $2.07$ billion image-retrieval requests for $92.3$ million distinct images. Thus, each image is requested about $22$ times on average. Accesses are highly skewed: the hottest $1\%$ of images account for $38\%$ of requests, and the hottest $10\%$ account for $71\%$ (\autoref{fig:civitai-pop}). 

This skew makes caching effective. An LRU cache sized to $10\%$ of the working set, or about $9.2$ million images, serves $87.2\%$ of requests because temporal locality keeps recently accessed images resident beyond what static popularity alone would predict. \autoref{fig:civitai-miss} shows the per-interval miss ratio over the full replay for caches sized at $1\%$, $5\%$, and $10\%$ of the working set; the dashed lines indicate the corresponding whole-trace miss ratios of $33.8\%$, $18.6\%$, and $12.8\%$.
The miss ratio increases over time as the corpus grows and a fixed working-set fraction covers a shrinking share of the absolute hot set. After around 1.5 years, the miss ratio starts to stablize. 

% This concentration makes a hot cache backed by on-demand regeneration of the cold tail attractive, and allows us to replay the trace directly against the caching threshold in \autoref{eq:cachethreshpow}, rather than relying on the idealized Zipfian model in \autoref{sec:caching}.

\begin{figure}[t]
  \centering
  \begin{subfigure}{0.47\columnwidth}
    \includegraphics[width=\textwidth]{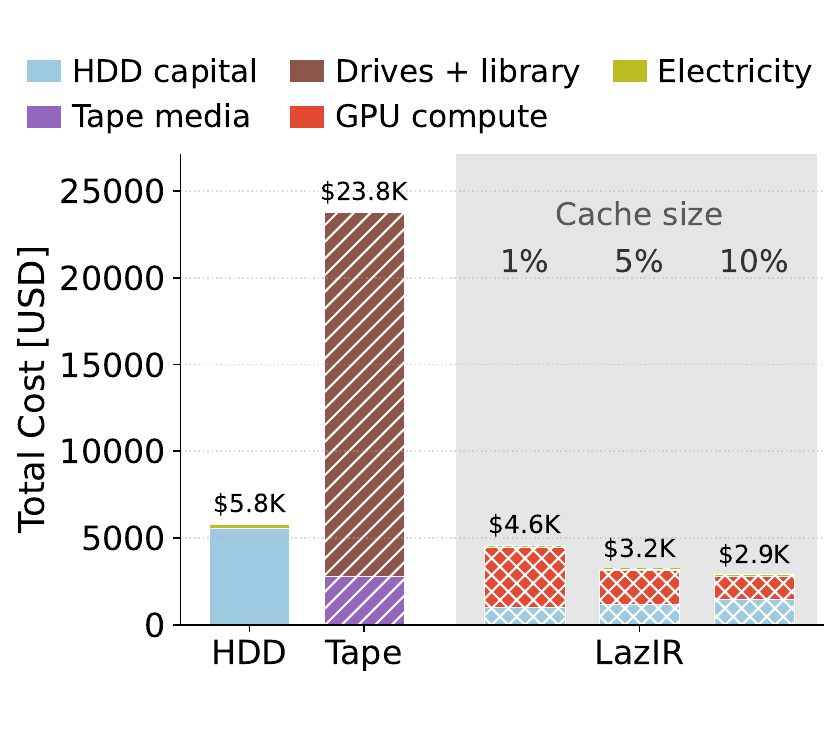}
    \vspace{-2em}
    \caption{\vspace{-1.0em}Storage and LazIR cost breakdown.}
    \label{fig:civitai-cost-stack}
  \end{subfigure}
  \hfill
  \begin{subfigure}{0.49\columnwidth}
    \raisebox{6.4pt}[\height][\depth]{%
      \includegraphics[width=\textwidth]{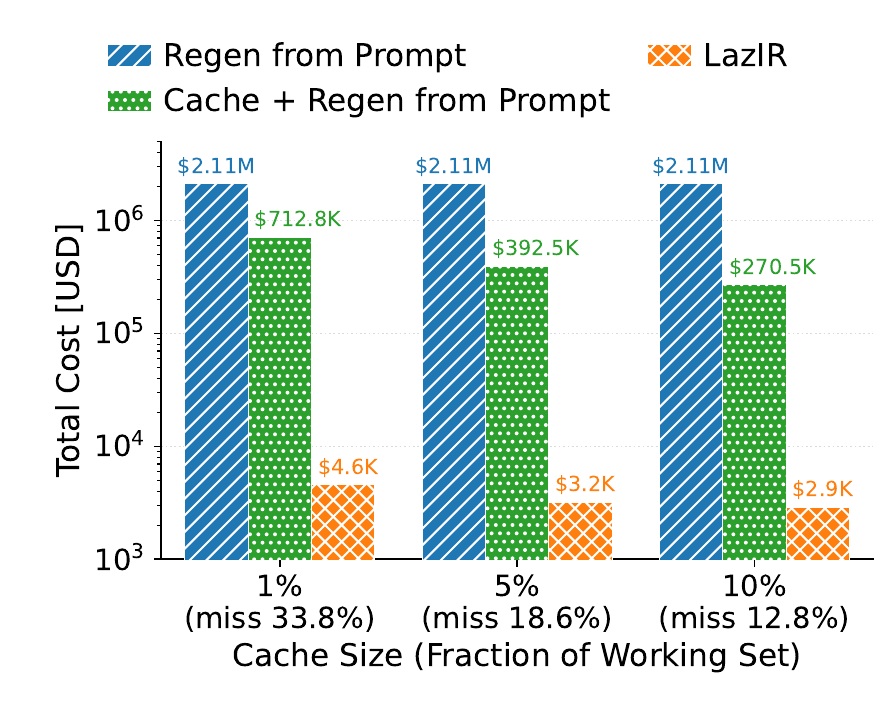}
      }
    \vspace{-2em}
    \caption{\vspace{-1.0em}Impact of cache size on regeneration.}
    \label{fig:civitai-cost-lines}
  \end{subfigure}
  \caption{Cost evaluation using the CompanyX trace replay at three LRU cache sizes. (a) Cost breakdown comparing materialized-artifact storage on HDD or tape with LazIR-based regeneration, with LazIR consistently achieving the lowest cost. (b) Prompt-based regeneration remains expensive even with caching, whereas LazIR reduces serving cost by roughly two orders of magnitude.}
  \vspace{-10pt}
  \label{fig:civitai-policy-cost}
\end{figure}

\subsection{Cost Analysis}
\label{sec:ir-cost}
Leveraging this production workload, we conduct a trace-driven simulation to compare storage and regeneration costs. We evaluate three regeneration policies---prompt-based regeneration, cached prompt-based regeneration, and LazIR---against two materialized-artifact storage baselines on HDD and tape, under 2025 prices using FLUX.1-dev model (\autoref{fig:civitai-policy-cost}).

\smallskip
\noindent\textbf{Materialized-artifact storage is around $100\times$ cheaper than prompt-based regeneration.}
On the trace replay, materialized-artifact storage on HDD is the cheapest conventional baseline:
persisting all $92.3$ million generated images costs about \$5,800 over the $2.87$-year trace. Materialized-artifact storage on tape costs about \$23,800. Although tape media is cheaper per byte than HDD, the corpus is only $0.28$ PB, so the fixed costs of tape drives and library slots are not yet amortized.
Prompt-based regeneration is far more expensive. Regenerating all $2.07$ billion requests with FLUX.1-dev costs about \$2.11 million, roughly $364\times$ more than materialized-artifact storage on HDD. Caching reduces the number of regenerations but does not change the conclusion. With a cache holding $10\%$ of the working set, the miss ratio drops to $12.8\%$, yet cached prompt-based regeneration still costs about \$270,000, or $47\times$ more than materialized-artifact storage on HDD, because each miss still reruns the expensive generation pipeline.

\smallskip
\noindent \textbf{LazIR reduces storage cost by 50\%.} Switching from prompt-based regeneration to intermediate representations (LazIR) reduces the cost to around \$2,900, 100$\times$ cheaper than prompt-based regeneration, and 50\% cheaper than materialized-artifact storage on HDD.
% It keeps the same hot cache, stores a compact latent for every image, and decodes cache misses from the latent, reducing total cost to about \$2,900. 
While caching alone reduces how often regeneration occurs, the key enabler of the cost reduction is that intermediate representation reduces the compute cost of each on-demand regeneration by 200$\times$ (\autoref{tab:modality-ex}). 
In the meantime, the intermediate representation is compact. 
Compared with materialized-artifact storage, which stores $276.8$ TB of materialized images, LazIR stores IR with a 1/6 footprint and a small cache for hot artifacts.

\smallskip
\noindent\textbf{Cache-size sensitivity.}
We assumed a 10\% cache size in the previous discussion, ~\autoref{fig:civitai-policy-cost} shows how regeneration cost changes with cache size. We observe that cache size increases only slightly increases the cost of the cache, while significantly reduces the cost of regeneration. For LazIR, increasing cache size from 1\% to 10\% reduces total cost by more than 36\%. 

% , while the cost of LazIR decreases with increasing cache size. 

% and ~\autoref{fig:civitai-cost-stack} breaks them down against the LazIR
% policy: full HDD storage stays at about \$5{,}800 and full tape at about
% \$23{,}800, tape being the more expensive of the two here because its fixed
% drive and library infrastructure dominates at this small $0.28$\,PB corpus,
% where a single pair of \$10{,}000 LTO drives alone outweighs the entire HDD
% capital.
% Full regeneration is likewise flat at \$2.11\,M regardless of cache size, since
% it holds nothing and regenerates every request.
% ~\autoref{fig:civitai-cost-lines} plots the three regeneration policies
% against cache size: cache plus full regeneration falls from \$713\,K at $1\%$ to
% \$270\,K at $10\%$ as fewer requests miss, yet it stays $47$ to $123\times$
% above full storage because each miss still pays the full $2{,}100$\,TFLOP
% generation cost.
% Cache plus IR cold store instead decodes each miss at $k_{\text{gen}} \text{flop}_{\text{regen,prompt}}{=}10.5$\,TFLOP, so
% its total drops from \$4{,}600 at $1\%$ to \$2{,}900 at $10\%$, staying below
% full HDD at every cache size and roughly $5\times$ below full tape.
% Two effects trade off as the cache grows, the hot cache and its power cost rise
% with size while miss compute shrinks, and because miss decode is already cheap
% under the IR store the net gain from a larger cache is modest, so the LazIR
% policy wins comfortably across the whole cache-size range.

\subsection{Serving Performance Analysis}
\label{sec:ir-perf}
Cost is only half the story, since the LazIR is viable only if the on-demand decoding is fast enough for interactive retrieval.

\smallskip
\noindent\textbf{Experiment Setup.}
We replay the same trace assuming an elastic GPU pool that decodes each miss from its latent (40~ms ~\autoref{tab:modality-ex}), and we measure the serving latency and the GPU count the pool provisions.
The pool is serverless with auto-scaling: it starts from a single GPU and adds one 
whenever the per-GPU backlog exceeds a target \textit{queue size}, and each added GPU pays a $150$ ms cold start before it begins serving. 
It releases capacity gradually rather than all at once, holding an idle GPU warm
for $1$\,s and shutting one down only once demand falls below the active GPU count, which prevents the pool from repeatedly paying the cold start as load fluctuates.
% The target queue size is the one knob that governs this trade, since it sets how
% much backlog the pool tolerates before it scales out, and we sweep it from a
% short queue that scales eagerly to a long queue that leans on buffering.

\begin{figure}[t]
  \centering
  \begin{subfigure}{0.49\columnwidth}
    \centering
    \includegraphics[width=\textwidth]{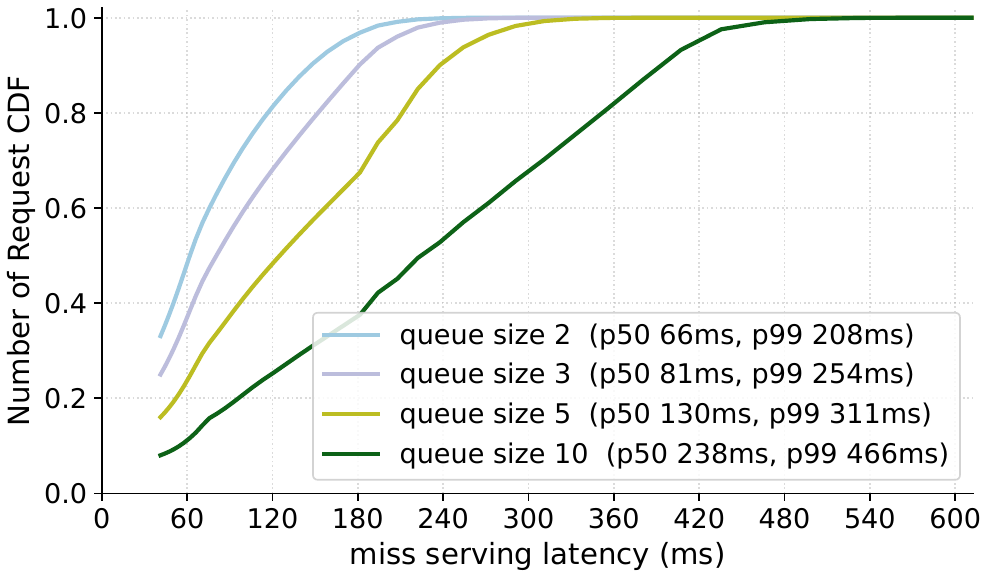}
    \caption{Miss serving-latency CDF by queue size.}
    \label{fig:civitai-serverless-cdf}
  \end{subfigure}
  \hfill
  \begin{subfigure}{0.49\columnwidth}
    \centering
    \includegraphics[width=\textwidth]{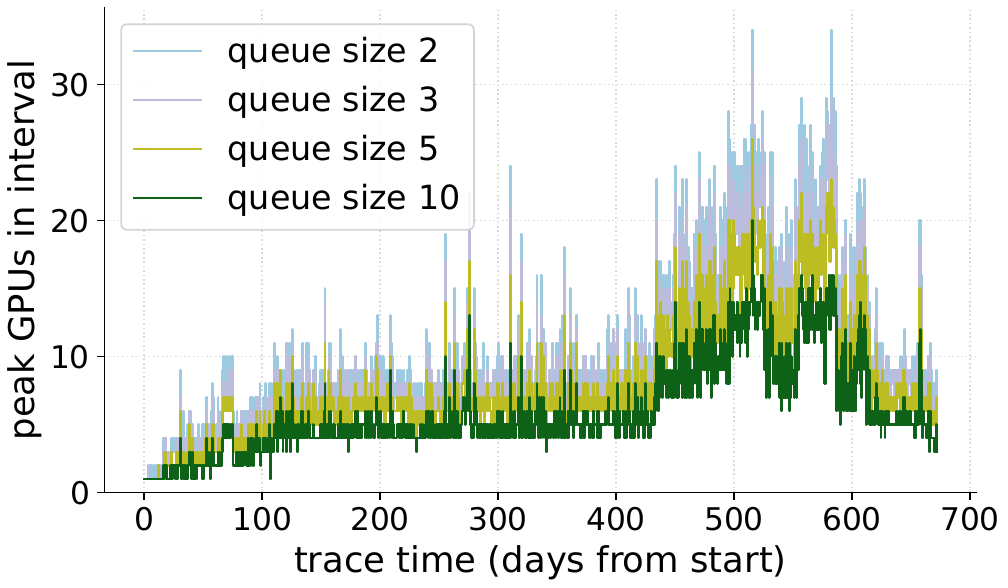}
    \caption{Peak GPU count over the trace by queue size.}
    \label{fig:civitai-serverless-gpu}
  \end{subfigure}
  \caption{Miss serving performance using an elastic GPU pool (150 ms cold-start). (a) Smaller queue size limit reduces serving latency, with median latency below 240 ms across all queue-size settings. (b) Requests are highly bursty, requiring additional GPU during peak load. Note that cache warmup period (no miss) is not shown. \vspace{-1.5em}}
  \label{fig:civitai-serverless}
\end{figure}

\smallskip
\noindent\textbf{LazIR provides interactive latency.}
\autoref{fig:civitai-serverless-cdf} shows that the miss-serving latency, including decoding and queueing, remains acceptable at both the median and the tail across different queue-size limits. With a queue-size limit of 2, LazIR achieves a median latency of 66 ms and a P99 latency of 208 ms. 
Note that these results report only miss latency. Since most requests are cache hits, the end-to-end serving latency across all requests is substantially lower. For example, with a 10\% miss ratio, a mean hit latency of 40 ms and miss latency of 200 ms, the mean overall latency will be 56 ms. 
For comparison, AWS reports that latency-sensitive applications using Amazon S3 can achieve a latency of roughly 100–200 ms~\cite{aws_s3_perf}. 

This low latency, however, comes at a cost. Unlike a smooth analytical model, the real-world trace exhibits substantial burstiness, which leads to spiky GPU allocation over time. \autoref{fig:civitai-serverless-gpu} shows the resulting peak GPU demand. A smaller queue-size limit reduces queueing latency, but it also increases the peak number of GPUs required to absorb request bursts. At peak, it requires 20-34 GPUs depending on the queue size. 

% \vspace{-15pt}
% \input{sections/carbon_emission}   % carbon emission commented out (kept, not deleted)
\section{Related Work}
\label{sec:related}

% \noindent\textbf{Energy and carbon of generative AI}
% A growing body of work measures the energy and carbon footprint of machine
% learning, from training-time accounting~\cite{patterson2021carbon,anthony2020carbontracker} to the inference cost of serving generative models at
% scale~\cite{luccioni2024power,ma2025inference}.
% These studies quantify the cost of producing an object once, whereas our
% concern is the recurring choice between paying that cost again on every access
% and amortising it through storage over a multi-year corpus.

\noindent\textbf{Cold storage and alternative media.}
Retaining large corpus cheaply has motivated the purpose-built photo stores~\cite{beaver2010finding,colarelli2002massive}, denser disk, and long-horizon archival media such as tape~\cite{lto2025roadmap,storer2008pergamum,legtchenko2016flamingo} and DNA~\cite{church2012nextgen,erlich2017dnafountain,organick2018random}.
This line treats the artifact as something that must be preserved and optimizes the medium on which it rests. Instead, we ask whether keeping the artifact at all is the cheaper option once its generator can reproduce it, and we use tape as one materialized-artifact storage baseline in that comparison.

% \smallskip
\noindent\textbf{Neural compression and intermediate representations.}
Learned image compression~\cite{cheng2020learned} and the latent spaces of modern generators~\cite{kingma2013auto,rombach2022ldm} show that an object can be regenerated from a representation far smaller than itself. Our LazIR design reuses this property but frames the latent not as a codec output but as a cold-store entry, whose size fraction $k_{\text{size}}$ and decode cost fraction $k_{\text{gen}}$ set an economic break-even against materialized-artifact storage.

% \smallskip
\noindent\textbf{Recomputing derived data instead of storing it.} The idea that derived data can be recomputed rather than stored recurs across communities: materialized-view selection in databases~\cite{harinarayan1996implementing,gupta1997selection}, regenerating cold derived datasets by re-executing their generating programs~\cite{gunda2010nectar}, lineage-based recovery of lost partitions~\cite{zaharia2012resilient,li2014tachyon}, and activation rematerialization in DNN training~\cite{chen2016training,jain2020checkmate}. In these settings, which intermediate state to retain is a per-workload tradeoff with no inherent winner. Generative pipelines differ in their cost structure: nearly all compute is front-loaded before the final decode, so the latent captures over 99\% of the generation cost at a fraction of the artifact's size. This structural asymmetry is what moves the storage-versus-regeneration break-even from decades away to today.

\section{Assumptions}
\label{sec:discussion}

% Although our analysis shows the intermediate representation usually
% dominates both storing pixels and regenerating from scratch, systems today
% rarely persist it, because storage abstractions are built around
% self-describing objects that decode with a stable, versioned codec, whereas
% a latent is only meaningful paired with the specific model that produced it.
% Adopting the latent as the storage unit thus trades a favourable cost
% position against a new dependence on model lifecycle, and the discussion
% below assumes this shift while examining what representation to store and
% what questions it leaves open.
%
% % \jason{add assumptions}
%

% \noindent\textbf{Assumptions.}
This work relies on several assumptions. First, we assume that current cost and technology trends persist over the next four decades. Exogenous shocks, however, can change these trajectories, as shown by the 2011 Thailand floods that disrupted the HDD supply chain and increased hard-drive prices~\cite{schneider2012thailandFlood}, and by recent AI-driven demand for HBM and high-capacity SSDs that has put upward pressure on memory and storage prices~\cite{trendforce2024hbm,trendforce2024aissd}. Second, we do not model inflation, which could modestly shift the break-even point. Third, we assume a skewed access pattern, which allows caching to reduce regeneration cost; other access distributions would change the effectiveness of caching and shift the break-even point. Fourth, we do not consider compression of generated artifacts or intermediate representations, either of which could affect the storage-regeneration tradeoff. Finally, we assume that regeneration reproduces the exact original artifact, although differences in batch size, GPU model, software stack, or numerical nondeterminism may introduce small bit-level differences~\cite{pytorchReproducibility,huggingfaceDiffusersReproducibility}.

\section{Conclusion}
\label{sec:conclusion}

AI-generated content is accumulating rapidly, while GPU compute cost continues to fall. This divergence makes the choice between storing generated artifacts and regenerating them on demand an increasingly important economic decision. This paper develops a lifecycle cost model to quantify this tradeoff. 
% The model jointly captures corpus growth, HDD and tape price and energy trends, drive replacement, request skew, caching, and GPU price-performance improvements. 
Our analysis shows that prompt-based regeneration is two orders of magnitude more expensive than materialized-artifact storage. It does not break even until the mid-2040s. 
The structure of generative pipelines changes this conclusion. Nearly all compute happens before a lightweight decode. A compact intermediate representation, therefore, captures the expensive work at a fraction of the artifact's size. 
LazIR caches hot objects, persists generated content as intermediate representations, and decodes on miss. This design moves the break-even from decades away to today. It cuts costs by $2.6\times$ to $10.5\times$ across image and music workloads. Evaluations on a production trace confirm these findings. As generative models become a dominant source of digital content, future storage systems should prioritize IR as the unit of long-term retention.

\bibliographystyle{ACM-Reference-Format}
\bibliography{reference}

@misc{adobeimage,
  author       = {{Adobe}},
  title        = {Adobe Firefly Delivers Creator-Focused, Commercially Viable Generative AI to Millions},
  year         = {2024},
  howpublished = {\url{https://blog.adobe.com/en/publish/2025/04/24/adobe-firefly-next-evolution-creative-ai-is-here}},
  note         = {Accessed April 28, 2026}
}

@misc{epochAI,
  author       = {{Epoch AI}},
  title        = {Trends in Machine Learning Hardware: GPU Price--Performance},
  year         = {2024},
  howpublished = {\url{https://epoch.ai/data/machine-learning-hardware}},
  note         = {Accessed: 2026-04-28}
}

@inproceedings{copet2023musicgen,
  author    = {Jade Copet and Felix Kreuk and Itai Gat and Tal Remez and David Kant and Gabriel Synnaeve and Yossi Adi and Alexandre D{\'e}fossez},
  title     = {Simple and Controllable Music Generation},
  booktitle = {Advances in Neural Information Processing Systems (NeurIPS)},
  year      = {2023},
  url       = {https://arxiv.org/abs/2306.05284},
  note      = {MusicGen: single-stage autoregressive transformer over
               EnCodec audio tokens; 3.3B-parameter Large variant},
  publisher = {arxiv}
}

@article{defossez2022encodec,
  author    = {Alexandre D{\'e}fossez and Jade Copet and Gabriel Synnaeve and Yossi Adi},
  title     = {High Fidelity Neural Audio Compression},
  journal   = {Transactions on Machine Learning Research (TMLR)},
  year      = {2023},
  url       = {https://arxiv.org/abs/2210.13438},
  note      = {EnCodec: neural audio codec with residual vector-quantization
               codebooks; the token IR used by MusicGen},
}

@misc{ramesh2022dalle2,
  author       = {Aditya Ramesh and Prafulla Dhariwal and Alex Nichol and Casey Chu and Mark Chen},
  title        = {Hierarchical Text-Conditional Image Generation with {CLIP} Latents},
  year         = {2022},
  eprint       = {2204.06125},
  archivePrefix = {arXiv},
  primaryClass  = {cs.CV},
  url          = {https://arxiv.org/abs/2204.06125},
  note         = {DALL-E 2 technical report; public service launched April 2022},
}

@misc{midjourney2022,
  author       = {{Midjourney, Inc.}},
  title        = {Midjourney},
  year         = {2022},
  howpublished = {\url{https://www.midjourney.com/}},
  note         = {Text-to-image generation service; open beta launched July 2022},
}

@inproceedings{cheng2020learned,
  author    = {Zhengxue Cheng and Heming Sun and Masaru Takeuchi and Jiro Katto},
  title     = {Learned Image Compression with Discretized {Gaussian} Mixture Likelihoods and Attention Modules},
  booktitle = {Proceedings of the IEEE/CVF Conference on Computer Vision and Pattern Recognition (CVPR)},
  year      = {2020},
  url       = {https://arxiv.org/abs/2001.01568},
}

@article{erlich2017dnafountain,
  author    = {Yaniv Erlich and Dina Zielinski},
  title     = {{DNA} Fountain enables a robust and efficient storage architecture},
  journal   = {Science},
  volume    = {355},
  number    = {6328},
  pages     = {950--954},
  year      = {2017},
  doi       = {10.1126/science.aaj2038},
  url       = {https://www.science.org/doi/10.1126/science.aaj2038},
}

@article{church2012nextgen,
  author    = {George M. Church and Yuan Gao and Sriram Kosuri},
  title     = {Next-Generation Digital Information Storage in {DNA}},
  journal   = {Science},
  volume    = {337},
  number    = {6102},
  pages     = {1628},
  year      = {2012},
  doi       = {10.1126/science.1226355},
  url       = {https://www.science.org/doi/10.1126/science.1226355},
}

@misc{aws_s3_perf,
  title        = {{Best practices design patterns: optimizing Amazon S3 performance}},
  author       = {{Amazon Web Services}},
  howpublished = {\url{https://docs.aws.amazon.com/AmazonS3/latest/userguide/optimizing-performance.html}},
  note         = {Accessed: 2026-07-10}
}

@article{schneider2012thailandFlood,
  title        = {The Lessons of Thailand's Flood},
  author       = {David Schneider},
  journal      = {IEEE Spectrum},
  year         = {2012},
  month        = {October},
  url          = {https://spectrum.ieee.org/the-lessons-of-thailands-flood}
}

@misc{trendforce2024hbm,
  title        = {{HBM} Prices to Increase by 5--10\% in 2025, Accounting for Over 30\% of Total {DRAM} Value, Says {TrendForce}},
  author       = {{TrendForce}},
  year         = {2024},
  month        = {May},
  url          = {https://www.trendforce.com/presscenter/news/20240506-12125.html}
}

@misc{trendforce2024aissd,
  title        = {{AI} Demand Drives Rapid Growth in {QLC} Enterprise {SSD}; Enterprise {SSD} Contract Prices Expected to Rise},
  author       = {{TrendForce}},
  year         = {2024},
  month        = {April},
  url          = {https://www.trendforce.com/presscenter/news/20240423-12122.html}
}

@misc{pytorchReproducibility,
  title        = {Reproducibility},
  author       = {{PyTorch}},
  year         = {2026},
  url          = {https://docs.pytorch.org/docs/main/notes/randomness.html}
}

@misc{huggingfaceDiffusersReproducibility,
  title        = {Reproducible Pipelines},
  author       = {{Hugging Face}},
  year         = {2025},
  url          = {https://huggingface.co/docs/diffusers/using-diffusers/reusing_seeds}
}

@misc{eia_electricity,
  author    = {{U.S. Energy Information Administration}},
  title     = {Electric Power Monthly: Average Retail Price of Electricity},
  year      = {2024},
  howpublished = {Online},
  url       = {https://www.eia.gov/electricity/monthly/},
}

@misc{everypixel2024aistats,
  author    = {{Everypixel Journal}},
  title     = {People Are Creating an Average of 34 Million Images Per Day: Statistics for 2024},
  year      = {2024},
  howpublished = {Online},
  url       = {https://journal.everypixel.com/ai-image-statistics},
  note      = {Comprehensive analysis: cumulative $>$15 billion AI-generated images since 2022--2023; current production rate $\sim$34 million/day (12.4 billion/year). Breakdown: Stable Diffusion 80\% (12.59B), Midjourney 964M, DALL-E 2 916M, Adobe Firefly 1B+. Source: \url{https://journal.everypixel.com/ai-image-statistics}},
}

@misc{openai2025gpumelting,
  author    = {{OpenAI} and Sam Altman},
  title     = {ChatGPT Image Generation: ``Our GPUs Are Melting''},
  year      = {2025},
  howpublished = {Online},
  url       = {https://x.com/sama/status/1905296867145154688},
  note      = {More than 130 million users generated over 700 million images in the first week of GPT-4o native image generation; OpenAI imposed temporary rate limits, stating that its GPUs were ``melting'' under demand. See also \url{https://en.wikipedia.org/wiki/GPT_Image}},
}

@inproceedings{beaver2010finding,
  author    = {Doug Beaver and Sanjeev Kumar and Harry C. Li and
               Jason Sobel and Peter Vajgel},
  title     = {{Finding a Needle in Haystack: {Facebook}'s Photo Storage}},
  booktitle = {9th {USENIX} Symposium on Operating Systems Design and
               Implementation ({OSDI})},
  year      = {2010},
  pages     = {47--60},
  url       = {https://www.usenix.org/legacy/event/osdi10/tech/full_papers/Beaver.pdf},
  note      = {Describes conventional pixel-level photo storage at scale,
               motivating latent-first alternatives},
  publisher = {USENIX}
}

@inproceedings{rombach2022high,
  author    = {Robin Rombach and Andreas Blattmann and Dominik Lorenz and
               Patrick Esser and Bj{\"o}rn Ommer},
  title     = {{High-Resolution Image Synthesis with Latent Diffusion Models}},
  booktitle = {IEEE/CVF Conference on Computer Vision and Pattern
               Recognition ({CVPR})},
  year      = {2022},
  pages     = {10684--10695},
  doi       = {10.1109/CVPR52688.2022.01042},
  url       = {https://arxiv.org/abs/2112.10752},
}

@misc{podell2023sdxl,
  author        = {Dustin Podell and Zion English and Kyle Lacey and
                   Andreas Blattmann and Tim Dockhorn and Jonas M{\"u}ller and
                   Joe Penna and Robin Rombach},
  title         = {{SDXL}: Improving Latent Diffusion Models for
                   High-Resolution Image Synthesis},
  year          = {2023},
  eprint        = {2307.01952},
  archivePrefix = {arXiv},
  primaryClass  = {cs.CV},
  doi           = {10.48550/arXiv.2307.01952},
  url           = {https://arxiv.org/abs/2307.01952},
}

@inproceedings{xie2025sana,
  author    = {Enze Xie and Junsong Chen and Junyu Chen and Han Cai and
               Haotian Tang and Yujun Lin and Zhekai Zhang and Muyang Li and
               Ligeng Zhu and Yao Lu and Song Han},
  title     = {{SANA}: Efficient High-Resolution Image Synthesis with
               Linear Diffusion Transformers},
  booktitle = {International Conference on Learning Representations ({ICLR})},
  year      = {2025},
  url       = {https://openreview.net/forum?id=N8Oj1XhtYZ},
}

@misc{labs2025flux1kontextflowmatching,
  author       = {{Black Forest Labs}},
  title        = {{FLUX.1: A Family of Flow Matching Models for
                  Text-to-Image Generation}},
  year         = {2024},
  howpublished = {\url{https://blackforestlabs.ai/announcing-black-forest-labs/}},
  note         = {Introduces the FLUX.1 model family using rectified flow
                  matching. Model weights and code at
                  \url{https://github.com/black-forest-labs/flux}.
                  Accessed 2025-01}
}

@inproceedings{kingma2013auto,
  author    = {Diederik P. Kingma and Max Welling},
  title     = {{Auto-Encoding Variational Bayes}},
  booktitle = {International Conference on Learning Representations ({ICLR})},
  year      = {2014},
  eprint    = {1312.6114},
  archivePrefix = {arXiv},
  url       = {https://arxiv.org/abs/1312.6114},
  note      = {Foundation of the VAE decoder used in latent diffusion models.
               DOI: 10.48550/arXiv.1312.6114}
}

@misc{mccallum2023diskprice,
  author       = {John C. McCallum},
  title        = {Disk Drive Prices (1955--2023)},
  year         = {2023},
  howpublished = {\url{https://jcmit.net/diskprice.htm}},
  note         = {Monthly survey of consumer HDD prices from NewEgg.com,
                  1955--2023. Archived at
                  \url{https://web.archive.org/web/2024/https://jcmit.net/diskprice.htm}.
                  Accessed 2025-01},
}

@misc{wikipedia_hdd,
  author       = {{Wikipedia contributors}},
  title        = {Hard disk drive --- {Wikipedia, The Free Encyclopedia}},
  year         = {2025},
  howpublished = {\url{https://en.wikipedia.org/wiki/Hard_disk_drive}},
  note         = {Accessed 2025-02},
}

@misc{diskprices2025,
  author       = {{DiskPrices.com}},
  title        = {Disk Prices --- Current Hard Drive Cost per Gigabyte},
  year         = {2025},
  howpublished = {\url{https://diskprices.com/}},
  note         = {Accessed 2025-02},
}

@misc{epochai2024mlhardware,
  author       = {{Epoch AI}},
  title        = {Data on Machine Learning Hardware},
  year         = {2024},
  howpublished = {\url{https://epoch.ai/data/machine-learning-hardware}},
  note         = {Dataset of $>$170 AI accelerators (GPUs, TPUs) with
                  performance, price, and efficiency metrics.
                  CC-BY 4.0. Accessed 2026-02},
}

@misc{epa_egrid2023,
  author       = {{U.S. Environmental Protection Agency}},
  title        = {{Emissions \& Generation Resource Integrated Database
                  (eGRID) 2022}},
  year         = {2024},
  howpublished = {\url{https://www.epa.gov/egrid}},
  note         = {eGRID2022 national average grid emission factor:
                  $\approx$0.386\,kg\,CO$_2$/kWh (non-baseload
                  0.434\,kg\,CO$_2$/kWh). Accessed 2026-02},
}

@misc{google2024environmental,
  author       = {{Google LLC}},
  title        = {{Google 2024 Environmental Report}},
  year         = {2024},
  howpublished = {\url{https://sustainability.google/reports/
                  google-2024-environmental-report/}},
  note         = {Reports 100\% renewable energy matching and
                  market-based scope\,2 emissions near zero;
                  location-based intensity for US data centres
                  0.10--0.20\,kg\,CO$_2$/kWh. Accessed 2026-02},
}

@misc{microsoft2024sustainability,
  author       = {{Microsoft Corporation}},
  title        = {{Microsoft 2024 Environmental Sustainability Report}},
  year         = {2024},
  howpublished = {\url{https://query.prod.cms.rt.microsoft.com/cms/api/am/binary/RW1lGvn}},
  note         = {Discloses data-centre PPA portfolio achieving
                  effective carbon intensity below 0.20\,kg\,CO$_2$/kWh
                  across US regions. Accessed 2026-02},
}

@techreport{iea2025energyai,
  author      = {{International Energy Agency (IEA)}},
  title       = {{Energy and AI: World Energy Outlook Special Report}},
  institution = {International Energy Agency},
  year        = {2025},
  url         = {https://www.iea.org/data-and-statistics/charts/global-data-centre-co2-emissions-base-case-2020-2035},
  note         = {Base Case: data-centre electricity demand rises from
                 ${\sim}$270\,TWh (2020) to ${\sim}$415\,TWh (2024) and
                 ${\sim}$945\,TWh (2030); related CO$_2$ emissions grow from
                 ${\sim}$180\,Mt today toward a ${\sim}$320\,Mt peak by 2030,
                 then ease to ${\sim}$300\,Mt by 2035 as grid decarbonisation
                 outpaces demand growth. Report:
                 \url{https://www.iea.org/reports/energy-and-ai}.
                 Accessed 2026-06},
}

@misc{seagate_exos_datasheet,
  author       = {{Seagate Technology}},
  title        = {{Seagate EXOS 7E10 Hard Drive Data Sheet}},
  year         = {2022},
  howpublished = {\url{https://www.seagate.com/files/www-content/datasheets/pdfs/exos-7e10-DS1957-4-2107US-en_US.pdf}},
  note         = {EXOS 7E10: 5.0--5.3\,W typical active power at 10--16\,TB
                  capacity, yielding $\approx$0.31--0.50\,W/TB per drive.
                  System-level estimate (1\,W/TB) includes PUE overhead
                  and storage-networking equipment. Accessed 2026-02},
}

@misc{uptime_pue2023,
  author       = {{Uptime Institute}},
  title        = {{Global Data Center Survey Results 2023}},
  year         = {2023},
  howpublished = {\url{https://uptimeinstitute.com/resources/research-and-reports/uptime-institute-global-data-center-survey-results-2023}},
  note         = {Reports an average industry PUE of 1.58 globally in 2023;
                  large cloud operators average $\approx$1.40.
                  We adopt PUE\,1.4 as a conservative mid-range estimate.
                  Accessed 2026-02},
}

@misc{nvidia_b200_datasheet,
  author       = {{NVIDIA Corporation}},
  title        = {{NVIDIA Blackwell {B200} Tensor Core {GPU} / {DGX B200}
                  Data Sheet}},
  year         = {2024},
  howpublished = {\url{https://resources.nvidia.com/en-us-dgx-systems/dgx-b200-datasheet}},
  note         = {B200 SXM: 2250\,TFLOP/s dense tensor-FP16, 1000\,W TDP,
                  yielding $\approx$0.44\,J/TFLOP as the 2025 anchor.
                  Accessed 2026-07},
}

@misc{seagate_exosm_datasheet,
  author       = {{Seagate Technology}},
  title        = {{Seagate EXOS M 30\,TB (Mozaic 3+) Hard Drive Data Sheet}},
  year         = {2025},
  howpublished = {\url{https://www.seagate.com/content/dam/seagate/en/content-fragments/products/datasheets/exos-m-v2/exos-m-v2-DS2045-3-2506-en_US.pdf}},
  note         = {EXOS M 30\,TB (ST30000NM004K): 6.9\,W idle, 9.5\,W max
                  operating (random read), yielding $\approx$0.23\,W/TB idle
                  and $\approx$0.32\,W/TB active per drive (bare-drive, 2025
                  anchor). Accessed 2026-07},
}

@misc{seagate2023lca,
  author       = {{Seagate Technology Holdings plc}},
  title        = {{Seagate Fiscal Year 2023 ESG Report and Product
                  Carbon Footprint Data}},
  year         = {2023},
  howpublished = {\url{https://www.seagate.com/global-citizenship/sustainability/}},
  note         = {Published product carbon footprint (PCF) for 3.5-inch
                  HDD product lines; embodied CO$_2$e values of
                  $\approx$5--20\,kg per drive depending on capacity
                  (covering materials, fabrication, and transport).
                  Accessed 2026-02},
}

@misc{wd2023lca,
  author       = {{Western Digital Corporation}},
  title        = {{Western Digital 2023 ESG Report: Product Environmental Data}},
  year         = {2023},
  howpublished = {\url{https://www.westerndigital.com/company/corporate-responsibility}},
  note         = {Life-cycle assessment (LCA) data for mechanical HDDs;
                  manufacturing-phase carbon footprint broadly consistent
                  with Seagate PCF disclosures. Accessed 2026-02},
}

@misc{nvidia2023sustainability,
  author       = {{NVIDIA Corporation}},
  title        = {{NVIDIA Corporate Responsibility Report 2023}},
  year         = {2023},
  howpublished = {\url{https://images.nvidia.com/aem-dam/Solutions/documents/FY2023-NVIDIA-Corporate-Responsibility-Report.pdf}},
  note         = {Scope~3 embodied-carbon data for A100-class datacenter GPUs;
                  device-level carbon footprint estimates correspond to
                  $\approx$0.50\,kg\,CO$_2$e/(TFLOP/s) at 312\,TFLOP/s FP32.
                  Accessed 2026-02},
}

@misc{lto2025roadmap,
  author       = {{LTO Program}},
  title        = {{LTO Program Announces New 40\,TB LTO-10 Cartridge
                  Specifications and Refreshes Its Roadmap}},
  year         = {2025},
  howpublished = {\url{https://www.lto.org/2025/11/lto-program-announces-new-40-tb-lto-10-cartridge-specifications-and-refreshes-its-roadmap-for-ultra-high-density-ai-ready-archival-storage/}},
  note         = {LTO-10: 30\,TB native at launch (June 2025), 40\,TB native
                  spec shipping Q1 2026. Accessed 2026-06},
}

@misc{fujifilm2025lto10,
  author       = {{Fujifilm}},
  title        = {{Fujifilm Launches LTO Ultrium 10 Data Cartridge}},
  year         = {2025},
  howpublished = {\url{https://www.fujifilm.com/us/en/news/data-storage/fujifilm-launches-lto-ultrium-10-data-cartridge}},
  note         = {30\,TB native (75\,TB at 2.5:1 compression); 40\,TB variant
                  available January 2026. Accessed 2026-06},
}

@misc{ltolifespan2024,
  author       = {{TechTarget (Brien Posey)}},
  title        = {{How to Estimate the Lifespan of LTO Tapes}},
  year         = {2024},
  howpublished = {\url{https://www.techtarget.com/searchdatabackup/tip/How-to-estimate-the-lifespan-of-LTO-tapes}},
  note         = {LTO media rated 15--30\,years under controlled storage;
                  best-practice data refresh every 10--20\,years. Accessed 2026-06},
}

@misc{techradar2025lto10,
  author       = {{TechRadar (Wayne Williams)}},
  title        = {{Tape Is Not Dead Yet, but \$300 LTO-10 Cartridges Won't
                  Help Its Cause}},
  year         = {2025},
  howpublished = {\url{https://www.techradar.com/pro/tape-is-not-dead-yet-but-usd300-lto-10-cartridges-and-inflated-exabyte-numbers-wont-help-its-cause}},
  note         = {LTO-10 cartridges priced \$275--315, roughly \$10/TB native,
                  about 2$\times$ the cheapest prior-generation cost per TB.
                  Accessed 2026-06},
}

@misc{tapeandmedia_ltoprices,
  author       = {{TapeAndMedia.com}},
  title        = {{LTO Ultrium Data Cartridge Retail Prices, LTO-6 through LTO-10}},
  year         = {2025},
  howpublished = {Archived product listings, Internet Archive Wayback Machine},
  note         = {Launch-era single-cartridge retail prices captured near each
                  generation's release, divided by native capacity: LTO-6
                  \$117.32/2.5\,TB (2013-01), LTO-7 \$186.68/6\,TB (2015-12),
                  LTO-8 \$188.00/12\,TB (2017-12), LTO-9 \$202.50/18\,TB
                  (2021-09), LTO-10 \$312.16/30\,TB (2025-06); native \$/TB
                  46.9, 31.1, 15.7, 11.3, 10.4. Dated snapshots:
                  
                  Accessed 2026-07},
}

@misc{spectra2025lto10drive,
  author       = {{Spectra Logic}},
  title        = {{LTO-10 Tape Drive Data Sheet}},
  year         = {2025},
  howpublished = {\url{https://spectralogic.com/resources/lto-10-tape-drive-data-sheet/}},
  note         = {LTO-10 native transfer rate 400\,MB/s (1000\,MB/s
                  compressed); standalone LTO-10 drives priced over \$10{,}000.
                  Accessed 2026-06},
}

@misc{ltodrivelife2024,
  author       = {{Media Duplication Systems}},
  title        = {{Understanding LTO Tape Life Expectancy: Key Factors and Insights}},
  year         = {2024},
  howpublished = {\url{https://www.mediaduplicationsystems.com/blog/understanding-lto-tape-life-expectancy-key-factors-and-insights/}},
  note         = {LTO drive MTBF ~250{,}000 hours; recommended tape-drive
                  replacement roughly every 4--5 years in practice. Accessed 2026-07},
}

@inproceedings{jeon2019analysis,
  author       = {Myeongjae Jeon and Shivaram Venkataraman and Amar Phanishayee
                  and Junjie Qian and Wencong Xiao and Fan Yang},
  title        = {Analysis of Large-Scale Multi-Tenant {GPU} Clusters for {DNN}
                  Training Workloads},
  booktitle    = {2019 USENIX Annual Technical Conference (USENIX ATC 19)},
  pages        = {947--960},
  year         = {2019},
  note         = {Production Microsoft ``Philly'' trace: average GPU utilization
                  is low (large distributed jobs average ~40\%), motivating a
                  fractional GPU duty anchor. Accessed 2026-07},
}

@misc{tapelibrary2024,
  author       = {{TechTarget}},
  title        = {{Buying a Backup Tape Library: What You Need to Know}},
  year         = {2024},
  howpublished = {\url{https://searchsmbstorage.techtarget.com/tip/Buying-your-first-backup-tape-library-What-you-need-to-know}},
  note         = {Tape libraries start ~\$5{,}000 for tens of slots and scale to
                  enterprise frames of thousands of slots (e.g. Spectra T950 up
                  to 10{,}020 LTO slots); per-slot frame plus robotics cost is
                  estimated at ~\$100/slot at enterprise scale. Accessed 2026-06},
}

@misc{fujifilm2024tco,
  author       = {{Fujifilm Data Storage}},
  title        = {{Tape Storage vs.\ Disk Storage: Total Cost of Ownership}},
  year         = {2024},
  howpublished = {\url{https://datastorage-na.fujifilm.com/tape-storage-vs-disk-storage-getting-the-facts-straight-about-total-cost-of-ownership/}},
  note         = {Tape archive consumes ~87--96\% less energy than a comparable
                  disk array over 10 years, since idle cartridges draw no power.
                  Accessed 2026-06},
}

@misc{aiimagevolume2026,
  author       = {{Imagera}},
  title        = {{AI Image Generation Statistics 2026}},
  year         = {2026},
  howpublished = {\url{https://imagera.ai/blog/ai-image-generation-statistics-2026}},
  note         = {Aggregate AI image generation across all platforms grew from
                  about 34 million images per day in 2023 to roughly 80 million
                  per day by 2026. Source: \url{https://imagera.ai/blog/ai-image-generation-statistics-2026}. Accessed 2026-06},
}

@inproceedings{yang2020twitter,
  author    = {Juncheng Yang and Yao Yue and K. V. Rashmi},
  title     = {A Large Scale Analysis of Hundreds of In-memory Cache Clusters at {Twitter}},
  booktitle = {14th USENIX Symposium on Operating Systems Design and Implementation (OSDI 20)},
  year      = {2020},
  pages     = {191--208},
  publisher = {USENIX Association},
}

@inproceedings{breslau1999zipf,
  author    = {Lee Breslau and Pei Cao and Li Fan and Graham Phillips and Scott Shenker},
  title     = {Web Caching and {Zipf}-like Distributions: Evidence and Implications},
  booktitle = {Proceedings of IEEE INFOCOM '99},
  year      = {1999},
  volume    = {1},
  pages     = {126--134},
}

@article{harinarayan1996implementing,
  title={Implementing data cubes efficiently},
  author={Harinarayan, Venky and Rajaraman, Anand and Ullman, Jeffrey D},
  journal={Acm Sigmod Record},
  volume={25},
  number={2},
  pages={205--216},
  year={1996},
  publisher={ACM New York, NY, USA}
}

@inproceedings{gupta1997selection,
  title={Selection of views to materialize in a data warehouse},
  author={Gupta, Himanshu},
  booktitle={International Conference on Database Theory},
  pages={98--112},
  year={1997},
  publisher={Springer}
}

@inproceedings{gunda2010nectar,
  title={Nectar: automatic management of data and computation in datacenters},
  author={Gunda, Pradeep Kumar and Ravindranath, Lenin and Thekkath, Chandramohan A and Yu, Yuan and Zhuang, Li},
  booktitle={9th USENIX Symposium on Operating Systems Design and Implementation (OSDI 10)},
  year={2010}
}

@inproceedings{zaharia2012resilient,
  title={Resilient distributed datasets: A $\{$Fault-Tolerant$\}$ abstraction for $\{$In-Memory$\}$ cluster computing},
  author={Zaharia, Matei and Chowdhury, Mosharaf and Das, Tathagata and Dave, Ankur and Ma, Justin and McCauly, Murphy and Franklin, Michael J and Shenker, Scott and Stoica, Ion},
  booktitle={9th USENIX symposium on networked systems design and implementation (NSDI 12)},
  pages={15--28},
  year={2012}
}

@inproceedings{li2014tachyon,
  title={Tachyon: Reliable, memory speed storage for cluster computing frameworks},
  author={Li, Haoyuan and Ghodsi, Ali and Zaharia, Matei and Shenker, Scott and Stoica, Ion},
  booktitle={Proceedings of the ACM Symposium on Cloud Computing},
  publisher = {Association for Computing Machinery},
  pages={1--15},
  year={2014}
}

@article{chen2016training,
  title={Training deep nets with sublinear memory cost},
  author={Chen, Tianqi and Xu, Bing and Zhang, Chiyuan and Guestrin, Carlos},
  journal={arXiv preprint arXiv:1604.06174},
  year={2016}
}

@article{jain2020checkmate,
  title={Checkmate: Breaking the memory wall with optimal tensor rematerialization},
  author={Jain, Paras and Jain, Ajay and Nrusimha, Aniruddha and Gholami, Amir and Abbeel, Pieter and Gonzalez, Joseph and Keutzer, Kurt and Stoica, Ion},
  journal={Proceedings of Machine Learning and Systems},
  volume={2},
  pages={497--511},
  year={2020}
}

@inproceedings{ho2020ddpm,
  title     = {Denoising Diffusion Probabilistic Models},
  author    = {Ho, Jonathan and Jain, Ajay and Abbeel, Pieter},
  booktitle = {Advances in Neural Information Processing Systems},
  volume    = {33},
  pages     = {6840--6851},
  year      = {2020},
  publisher = {Curran Associates, Inc.}
}

@inproceedings{rombach2022ldm,
  title     = {High-Resolution Image Synthesis with Latent Diffusion Models},
  author    = {Rombach, Robin and Blattmann, Andreas and Lorenz, Dominik and Esser, Patrick and Ommer, Bj{\"o}rn},
  booktitle = {Proceedings of the IEEE/CVF Conference on Computer Vision and Pattern Recognition},
  pages     = {10684--10695},
  year      = {2022}
}

@inproceedings{saharia2022imagen,
  title     = {Photorealistic Text-to-Image Diffusion Models with Deep Language Understanding},
  author    = {Saharia, Chitwan and Chan, William and Saxena, Saurabh and Li, Lala and Whang, Jay and Denton, Emily and Seyed Ghasemipour, Seyed Kamyar and Ayan, Burcu Karagol and Mahdavi, S. Sara and Lopes, Rapha Gontijo and Salimans, Tim and Ho, Jonathan and Fleet, David J. and Norouzi, Mohammad},
  booktitle = {Advances in Neural Information Processing Systems},
  volume    = {35},
  pages     = {36479--36494},
  year      = {2022},
  publisher = {Curran Associates, Inc.}
}

@inproceedings{colarelli2002massive,
  title={Massive arrays of idle disks for storage archives},
  author={Colarelli, Dennis and Grunwald, Dirk},
  booktitle={SC'02: Proceedings of the 2002 ACM/IEEE Conference on Supercomputing},
  pages={47--47},
  year={2002},
  organization={IEEE}
}

@inproceedings{storer2008pergamum,
  title={Pergamum: Replacing Tape with Energy Efficient, Reliable, Disk-Based Archival Storage.},
  author={Storer, Mark W and Greenan, Kevin M and Miller, Ethan L and Voruganti, Kaladhar},
  booktitle={Fast},
  volume={8},
  pages={1--16},
  year={2008}
}

@inproceedings{legtchenko2016flamingo,
  title={Flamingo: Enabling Evolvable $\{$HDD-based$\}$$\{$Near-Line$\}$ Storage},
  author={Legtchenko, Sergey and Li, Xiaozhou and Rowstron, Antony and Donnelly, Austin and Black, Richard},
  booktitle={14th USENIX Conference on File and Storage Technologies (FAST 16)},
  pages={213--226},
  year={2016}
}

@article{organick2018random,
  title={Random access in large-scale DNA data storage},
  author={Organick, Lee and Ang, Siena Dumas and Chen, Yuan-Jyue and Lopez, Randolph and Yekhanin, Sergey and Makarychev, Konstantin and Racz, Miklos Z and Kamath, Govinda and Gopalan, Parikshit and Nguyen, Bichlien and others},
  journal={Nature biotechnology},
  volume={36},
  number={3},
  pages={242--248},
  year={2018},
  publisher={Nature Publishing Group}
}

@article{fagin1977asymptotic,
  title={Asymptotic miss ratios over independent references},
  author={Fagin, Ronald},
  journal={Journal of Computer and System Sciences},
  volume={14},
  number={2},
  pages={222--250},
  year={1977},
  publisher={Elsevier}
}

@inproceedings{fricker2012versatile,
  title={A versatile and accurate approximation for LRU cache performance},
  author={Fricker, Christine and Robert, Philippe and Roberts, James},
  booktitle={2012 24th international teletraffic congress (ITC 24)},
  pages={1--8},
  year={2012},
  organization={IEEE}
}

\newpage
\appendix
\section{Notation}
\label{app:notation}

Table\,\ref{tab:facts} and Table\,\ref{tab:notation} collect the constants and notation used across the three cost
models (Sections\,\ref{sec:hdd-cost-model}--\ref{sec:cache-ir}) and the
access-pattern derivation of Appendix\,\ref{app:access-pattern}, with each
symbol pointing to the equation that defines it. Table\,\ref{tab:facts} lists the
empirical parameter values, and Table\,\ref{tab:notation} defines the model notation.

\begin{table}[t]
  \centering
  \footnotesize
  \setlength{\tabcolsep}{3.5pt}
  \renewcommand{\arraystretch}{1.15}
  \caption{Empirically grounded parameters used in the cost models.}
  % The \emph{Energy and electricity} block lists the
  %          three per-unit electricity coefficients the model uses directly,
  %          $c_{\text{elec,HDD}}$, $c_{\text{elec,tape}}$, and
  %          $c_{\text{elec,GPU}}$, each derived in \autoref{sec:param-anchors} and
  %          \autoref{sec:regen-cost-model} from a device power density and the
  %          \$0.08/kWh rate. The HDD and GPU coefficients are 2025 anchors that
  %          decline at the same rate as the matching capital price, 10\%/yr for
  %          HDD and 30\%/yr for GPU, since the areal-density and throughput
  %          growth that cuts \$/PB and \$/TFLOP also cuts per-unit energy at the
  %          same pace.}
  \label{tab:facts}
  \begin{tabular}{@{}llll@{}}
    \toprule
    Symbol & Meaning & Value & Ref. \\
    \midrule
    \multicolumn{4}{@{}l}{\emph{Hardware price trends}}\\
    $c_{\text{HDD}}(2025)$ & HDD price anchor & 20{,}000 \$/PB & \cite{diskprices2025}\\
    $c_{\text{tape}}(2025)$ & tape price anchor (LTO-10) & 10{,}000 \$/PB & \cite{techradar2025lto10}\\
    $c_{\text{GPU}}(2025)$ & GPU price anchor & 18 \$/(TFLOPS) & \cite{epochai2024mlhardware}\\
    \midrule
    \multicolumn{4}{@{}l}{\emph{Device lifetime and duty}}\\
    $L$ & GPU lifetime & 3 yr & \cite{epochai2024mlhardware}\\
    $u$ & GPU utilization & 0.40 & \cite{jeon2019analysis}\\
    $L_{\text{HDD}}$ & HDD service life & 5 yr & \cite{seagate_exos_datasheet}\\
    $L_{\text{tape}}$ & tape media life & 30 yr & \cite{ltolifespan2024}\\
    $L_{\text{drive}}$ & tape drive life & 5 yr & \cite{ltodrivelife2024}\\
    \midrule
    \multicolumn{4}{@{}l}{\emph{Tape drive and library}}\\
    $c_{\text{drive}}$ & tape drive price & 10{,}000 \$/drive & \cite{spectra2025lto10drive}\\
    % $r_{\text{drive}}$ & drive transfer rate & 400 MB/s & \cite{spectra2025lto10drive}\\
    $c_{\text{slot}}$ & library slot cost & 100 \$/slot & \cite{tapelibrary2024}\\
    % $S_{\text{cart}}$ & cartridge capacity & 30 TB & \cite{fujifilm2025lto10}\\
    \midrule
    \multicolumn{4}{@{}l}{\emph{Energy and electricity}}\\
    $c_{\text{elec,HDD}}(2025)$ & HDD electricity coeff. & 315 \$/PB/yr & \cite{seagate_exosm_datasheet,uptime_pue2023,eia_electricity}\\
    $c_{\text{elec,tape}}(2025)$ & tape electricity coeff. & 31.5 \$/PB/yr & \cite{fujifilm2024tco,eia_electricity}\\
    $c_{\text{elec,GPU}}(2025)$ & GPU electricity coeff. & $9.78{\times}10^{-9}$ \$/TFLOP & \cite{nvidia_b200_datasheet,eia_electricity}\\
    % ---- carbon rows commented out (kept, not deleted) ----
    % \midrule
    % \multicolumn{4}{@{}l}{\emph{Carbon}}\\
    % $\gamma_e$ & grid carbon intensity & 0.30 kg/kWh & \cite{epa_egrid2023}\\
    % $p_C$ & carbon price & 50 \$/t & \cite{ecosystemmarketplace2023}\\
    % $\varepsilon_{\text{HDD}}$ & HDD embodied carbon & 5 kg/TB & \cite{seagate2023lca,wd2023lca}\\
    % $\varepsilon_{\text{GPU}}$ & GPU embodied carbon & 0.50 kg/(TF/s) & \cite{nvidia2023sustainability}\\
    \bottomrule
  \end{tabular}
\end{table}

\begin{table}[H]
  \centering
  \footnotesize
  \setlength{\tabcolsep}{3.5pt}
  \renewcommand{\arraystretch}{1.0}
  \caption{Notation used in the cost models of
           Sections\,\ref{sec:hdd-cost-model}--\ref{sec:cache-ir} and
           Appendix\,\ref{app:access-pattern}.
           Superscript $(g)$ marks the exponential-growth variant; the
           linear 1\,PB/yr baseline is its $g{=}0$ case.}
  \label{tab:notation}
  \begin{tabular}{@{}l p{0.62\linewidth}@{}}
    \toprule
    Symbol & Description \\
    \midrule
    \multicolumn{2}{@{}l}{\emph{Time and workload}}\\
    $t,\,y,\,\tau$ & calendar year; the study horizon is 2025--2045 \\
    $g$ & annual ingest growth rate \\
    $s_{\text{obj}}$ & size of one artifact [PB] \\
    $S_{\text{data}}(y)$ & new data ingested in year $y$, $(1+g)^{y-2025}$ [PB] \\
    $S^{\text{full}}_{\text{data}}(y)$ & cumulative live (powered) corpus volume [PB] \\
    $N(y)$ & cumulative artifact count through year $y$ \\
    $R^{(g)}(y)$ & annual request rate, $10\,N(y)$ (Eq.\,\ref{eq:nobjg}) \\
    \midrule
    \multicolumn{2}{@{}l}{\emph{Unit prices, lifetimes, and electricity}}\\
    $C_{\text{HDD}}(t)$, $C_{\text{tape}}(t)$ & HDD and tape price [\$/PB] (Eq.\,\ref{eq:phdd}; tape analogous, \S\ref{sec:tape-anchor}) \\
    $c_{\text{GPU}}(t)$ & GPU capacity price [\$/(TFLOP/s)] (Eq.\,\ref{eq:pgpu}) \\
    $\mathrm{drop}_{\text{HDD}}$, $\mathrm{drop}_{\text{tape}}$, $\mathrm{drop}_{\text{GPU}}$ & annual price decline rates \\
    $L$, $u$, $\Delta_{\text{yr}}$ & GPU service lifetime [yr], utilization, and seconds per year \\
    $c_{\text{TFLOP}}(t)$ & amortized GPU serving price [\$/TFLOP] (Eq.\,\ref{eq:ptflop}) \\
    $L_{\text{HDD}}$ & HDD service life before replacement [yr] \\
    $e$ & electricity rate [\$/kWh] \\
    $P^{\text{sys}}$ & system-level power density of a storage medium [W/PB] (Eq.\,\ref{eq:alpha}) \\
    $c_{\text{elec,HDD}}(t)$, $c_{\text{elec,tape}}(t)$ & HDD and tape electricity unit cost [\$/PB/yr] (Eqs.\,\ref{eq:alphahdd}, \ref{eq:alphatape}) \\
    $c_{\text{elec,GPU}}(t)$ & GPU electricity cost per unit compute [\$/TFLOP] \\
    \midrule
    \multicolumn{2}{@{}l}{\emph{Cost model 1: materialized-artifact storage}}\\
    $S_{\text{HDD,purchase}}(y)$ & HDD bought in year $y$: new ingest plus replacement [PB] (Eq.\,\ref{eq:provision}) \\
    $C_{\text{HDD,cum}}^{(g)}(y)$ & cumulative replacement-inclusive HDD capital (Eq.\,\ref{eq:chddcumrepl}) \\
    $C_{\text{tape,cum}}^{(g)}(y)$ & cumulative tape media cost (Eq.\,\ref{eq:ctapecumg}) \\
    $C_{\text{elec,HDD}}^{\text{full}}(y)$ & annual HDD electricity for materialized-artifact storage (Eq.\,\ref{eq:celecfull}) \\
    $E_{\text{full}}^{(g)}(y)$ & cumulative HDD electricity for materialized-artifact storage (Eq.\,\ref{eq:celecfullcum}) \\
    $C_{\text{total}}^{(g)}(y)$ & total materialized-artifact storage cost, capital plus electricity (Eq.\,\ref{eq:chddtotal}) \\
    \midrule
    \multicolumn{2}{@{}l}{\emph{Cost model 2: regeneration and caching}}\\
    $\text{flop}_{\text{regen,prompt}}$ & compute to generate one artifact from its prompt [TFLOP] \\
    $\text{flop}^{*}_{\text{regen,prompt}}$ & break-even generation cost below which a policy beats materialized-artifact storage [TFLOP] \\
    $C_{\text{regen}}^{(g)}(y)$ & annual compute cost of prompt-based regeneration (Eq.\,\ref{eq:cregeng}) \\
    $C_{\text{elec,regen}}(y)$ & annual GPU electricity of prompt-based regeneration (Eq.\,\ref{eq:celecregen}) \\
    $h$ & cache hit ratio; $H_M/H_N$ for the top $M$ of $N$ objects (Eq.\,\ref{eq:hitrate}) \\
    $F_{\text{cached}}$ & corpus fraction cached to reach $h$ (Eq.\,\ref{eq:cachesize}) \\
    $S_{\text{cache,data}}(y)$ & cache volume [PB] (Eq.\,\ref{eq:vcache}) \\
    $S_{\text{cache,purchase}}(y)$ & cache HDD capacity bought in year $y$ [PB] \\
    $C_{\text{cache,HDD}}(y)$ & annual HDD spend on the hot cache (Eq.\,\ref{eq:ccachehdd}) \\
    $C_{\text{miss}}(y)$ & annual regeneration cost of cache misses (Eq.\,\ref{eq:cmiss}) \\
    $C_{\text{mod}}(y)$ & annual cost of the cache policy (Eq.\,\ref{eq:cmod}) \\
    $C_{\text{elec,cache}}(y)$, $C_{\text{elec,GPU}}(y)$ & cache-HDD and miss-compute electricity (Eq.\,\ref{eq:celecgpu}) \\
    \midrule
    \multicolumn{2}{@{}l}{\emph{Cost model 3: LazIR intermediate-representation store}}\\
    $k_{\text{size}}$ & IR size as a fraction of the materialized artifact \\
    $k_{\text{gen}}$ & IR-regeneration compute as a fraction of prompt-based regeneration compute \\
    $S_{\text{IR,data}}(y)$ & IR (latent) store volume [PB] (Eq.\,\ref{eq:vir}) \\
    $C_{\text{IR,HDD}}(y)$ & annual HDD spend on the IR store \\
    $C_{\text{miss,IR}}(y)$ & annual decode-on-miss compute cost \\
    $C_{\text{elec,HDD}}^{\text{2tier}}(y)$, $C_{\text{elec,GPU}}^{\text{IR}}(y)$ & annual HDD and miss-decode GPU electricity (Eq.\,\ref{eq:celec2tier}) \\
    $C_{\text{mod}}^{(\text{IR})}(y)$ & annual cost of the LazIR policy (Eq.\,\ref{eq:cmodir}) \\
    $E_{\text{2tier}}(y)$ & cumulative electricity for HDD plus GPU in LazIR (Eq.\,\ref{eq:celec2tiercum}) \\
    \bottomrule
  \end{tabular}
\end{table}

\clearpage

\section{Access Pattern and Cache Fraction}
\label{app:access-pattern}

The caching policies of Section~\ref{sec:caching} are parameterized by the
cache fraction $F_{\text{cached}}$, the portion of the corpus that must be
held at full fidelity to serve a target fraction $h$ of all requests from the
cache.
This fraction is fixed entirely by how skewed the request popularity is.
Here we derive it in closed form under a Zipfian workload and report the
concrete values used in the main text.

We adopt the Zipfian distribution as a representative skewed workload for
user-facing generated content.
With Zipfian skew $s=1$, the probability of the $i$-th most popular object is
$p_i \propto 1/i$, so the share of requests served by the top $M$ of $N$
objects is
\begin{equation}
  h(M,N) = \frac{H_M}{H_N},
  \label{eq:hitrate}
\end{equation}
where $H_n = \sum_{i=1}^{n} 1/i \approx \ln n + \gamma$ with
$\gamma \approx 0.5772$ is the $n$-th harmonic number.
Setting $h(M,N) = h$ and using the harmonic approximation gives
$\ln M \approx h\ln N - (1-h)\,\gamma$, hence
\begin{equation}
  F_{\text{cached}} = \frac{M}{N} \approx e^{-(1-h)\gamma}\,N^{-(1-h)},
  \label{eq:cachesize}
\end{equation}
which decreases as the corpus grows.

\noindent\textbf{Result.}
At a target hit ratio $h = 0.90$, equivalently a $10\%$ miss ratio, this
reduces to $F_{\text{cached}} \approx 0.944\,N^{-0.10}$.
A corpus of $N = 10^{8}$ objects then needs only about $15\%$ of its objects
cached to serve $90\%$ of requests, and this fraction falls to about $9.4\%$
at $N = 10^{10}$.
A small hot set thus absorbs the vast majority of requests even as the total
catalog grows large, and the caching policies of Section~\ref{sec:caching}
use these values directly.

\section{Break-Even Walkthrough}
\label{sec:modality-walkthrough}

We evaluate every policy at the 2045 horizon under $g{=}20\%$ annual ingest growth and a
$h{=}0.90$ Zipfian hit ratio, counting both HDD capital with replacement and
the electricity to keep the corpus powered.
A policy is preferred over materialized-artifact storage on HDD when the model's generation cost
$\text{flop}_{\text{regen,prompt}}$ falls below that policy's break-even $\text{flop}^{*}_{\text{regen,prompt}}$, and every $\text{flop}^{*}_{\text{regen,prompt}}$ scales
linearly with $s_{\text{obj}}$.

\noindent\textbf{Storage cost, linear growth.}
Storing a constant 1\,PB of new objects each year on HDD over 2025--2045, with
drives re-bought every $L_{\text{HDD}}{=}5$ years, costs
$C_{\text{HDD,cum}}(2045){\approx}\$344{,}000$ in capital, up from the
\$178{,}000 single-purchase figure once replacement is included.
Keeping the cumulative corpus powered adds
$E_{\text{full}}(2045){\approx}\$21{,}000$ of electricity, for a
total linear-growth storage cost of $C_{\text{total}}(2045){\approx}\$365{,}000$.

\noindent\textbf{Storage cost, exponential growth.}
Under $g{=}20\%$ annual ingest growth the replacement-inclusive capital rises to
$C_{\text{HDD,cum}}^{(g)}(2045){\approx}\$1{,}515{,}300$ and electricity to
$E_{\text{full}}^{(g)}(2045){\approx}\$81{,}000$, giving a total
materialized-artifact storage baseline of $C_{\text{total}}^{(g)}(2045){\approx}\$1{,}597{,}000$ from
Equation\,\ref{eq:chddtotal}.
We use this exponential-growth total as the storage side of every break-even
below, the linear case being analogous at smaller magnitude.

\noindent\textbf{Break-even: prompt-based regeneration on every request.}
Prompt-based regeneration discards all materialized artifacts and incurs GPU
capital plus GPU electricity per object, so regeneration wins below
$\text{flop}^{*}_{\text{regen,prompt}}{=}C_{\text{total}}^{(g)}/\sum_\tau R^{(g)}(\tau)\bigl(c_{\text{TFLOP}}(\tau){+}c_{\text{elec,GPU}}(\tau)\bigr)$
from Equation\,\ref{eq:cthreshpow}.
Because the request rate $R^{(g)}$ scales as $1/s_{\text{obj}}$, the threshold scales
linearly with $s_{\text{obj}}$, giving $\text{flop}^{*}_{\text{regen,prompt}}{\approx}49.7$\,TFLOP for images and $84.4$\,TFLOP for
music.
Only MusicGen-Large at $21$\,TFLOP sits below its threshold, so prompt-based
regeneration beats materialized-artifact storage for music alone, whereas FLUX.1-dev at
$2{,}100$\,TFLOP favors materialized-artifact storage.

\noindent\textbf{Break-even: cached prompt-based regeneration.}
A Zipfian hot cache that serves $h{=}0.90$ of requests confines regeneration
to the $10\%$ that miss, at the price of the cache HDD capital and its
electricity, so caching wins below the threshold
$\text{flop}^{*}_{\text{regen,prompt}}$ of Equation\,\ref{eq:cachethreshpow}.
This lifts the thresholds to $\text{flop}^{*}_{\text{regen,prompt}}{\approx}454$\,TFLOP for images and $768$\,TFLOP for
music.
Caching pushes the music threshold well above MusicGen-Large's $21$\,TFLOP, but
FLUX.1-dev still exceeds its threshold, since regenerating a
$2{,}100$\,TFLOP image on every miss remains too costly
relative to its footprint.

\noindent\textbf{Break-even: Zipfian hot cache with IR cold store and regeneration on miss.}
The LazIR design stores a compact intermediate representation of every
object at fraction $k_{\text{size}}$ of its size and decodes a miss at reduced cost
$k_{\text{gen}} \text{flop}_{\text{regen,prompt}}$, so its threshold from
Equation\,\ref{eq:cachethreshirpow} is lifted by roughly
$(1{-}k_{\text{size}})/k_{\text{gen}}$ over the plain cache.
The thresholds climb to $\text{flop}^{*}_{\text{regen,prompt}}{\approx}74{,}200$\,TFLOP for images and
$359{,}000$\,TFLOP for music.
Both evaluated models now sit far below their thresholds: FLUX.1-dev at $2{,}100$\,TFLOP and
MusicGen-Large at $21$\,TFLOP, so once the latent is
stored and misses decode cheaply, regeneration on miss beats materialized-artifact
storage across both modalities.

\section{Carbon Emissions}
\label{sec:carbon}

Beyond the electricity cost in Section~\ref{sec:power-cost}, each kWh consumed
also produces CO$_2$ emissions, which increasingly matter for both social impact
and regulatory compliance. We therefore estimate cumulative lifecycle carbon
emissions for the full-HDD baseline and the two-tier system. We report emissions
in tonnes of CO$_2$ rather than converting them into a monetized carbon cost.

\smallskip
\noindent\textbf{Carbon coefficients.}
Carbon enters the model through four unit-emission coefficients. Two are
operational and scale with energy consumption: $\phi_{\mathrm{HDD}}$
[kg\,CO$_2$/PB/yr] for powered HDD storage and $\phi_{\mathrm{GPU}}$
[kg\,CO$_2$/TFLOP] for GPU compute. These are the carbon-intensity analogs of
the electricity-cost coefficients $\alpha$ from Section~\ref{sec:param-anchors}
and $\beta$ from Section~\ref{sec:recompute}. The other two are embodied and are
charged once per unit of manufactured hardware: $\Phi_{\mathrm{HDD}}$
[kg\,CO$_2$/PB] for HDD capacity and $\Psi_{\mathrm{GPU}}$ [kg\,CO$_2$/TFLOP]
for GPU compute capacity, with the latter amortized using the same lifetime and
utilization assumptions as $c_{\mathrm{TFLOP}}$.

These coefficients are derived from four device-level constants. We use a grid
carbon intensity of $\gamma_e = 0.30$\,kg\,CO$_2$/kWh for a typical US data
center. This is conservative relative to the 2023 US average of about
0.39\,kg\,CO$_2$/kWh~\cite{epa_egrid2023}, while large operators with renewable
power purchase agreements report values around 0.10--0.20\,kg\,CO$_2$/kWh
\cite{google2024environmental,microsoft2024sustainability}. For embodied
emissions, we use $\varepsilon_{\mathrm{HDD}} = 5$\,kg\,CO$_2$/TB for HDDs,
covering materials, fabrication, and transport
\cite{seagate2023lca,wd2023lca}. For GPUs, we use
$\varepsilon_{\mathrm{GPU}} = 0.50$\,kg\,CO$_2$/(TFLOP/s), corresponding to
roughly 150--200\,kg\,CO$_2$ per A100-class device at 312\,TFLOP/s dense
tensor-FP16 performance~\cite{nvidia2023sustainability}.

Reusing the HDD power density, GPU energy efficiency, and electricity price from
Section~\ref{sec:power-cost}, the operational carbon coefficients are
\begin{align}
  \phi_{\mathrm{HDD}}
    &= P_{\mathrm{HDD}}^{\mathrm{sys}} \times 8{,}760 \times
       \frac{\gamma_e}{10^{3}} \times \frac{10^3\,\mathrm{TB}}{\mathrm{PB}}
     = 2{,}628\;\mathrm{kg\,CO}_2/\mathrm{PB/yr},
  \label{eq:phihdd} \\
  \phi_{\mathrm{GPU}}
    &= \frac{\epsilon_{\mathrm{GPU}} \cdot \gamma_e}{3.6\times10^{6}}
     = 8.33\times10^{-8}\;\mathrm{kg\,CO}_2/\mathrm{TFLOP}.
  \label{eq:phigpu}
\end{align}
Thus, $\phi_{\mathrm{HDD}}/\alpha = \phi_{\mathrm{GPU}}/\beta =
\gamma_e/e = 3.75$\,kg\,CO$_2$/\$, meaning that operational carbon is exactly
$3.75\times$ the corresponding electricity cost in
Section~\ref{sec:power-cost}.

The embodied carbon coefficients are
\begin{align}
  \Phi_{\mathrm{HDD}}
    &= \varepsilon_{\mathrm{HDD}} \times 10^{3}\,\mathrm{TB/PB}
     = 5{,}000\;\mathrm{kg\,CO}_2/\mathrm{PB},
  \label{eq:Phihdd} \\
  \Psi_{\mathrm{GPU}}
    &= \frac{\varepsilon_{\mathrm{GPU}}}{L \cdot u}
     = \frac{0.50}{3.78\times10^{7}}
     = 1.32\times10^{-8}\;\mathrm{kg\,CO}_2/\mathrm{TFLOP}.
  \label{eq:Psigpu}
\end{align}
Since $\Psi_{\mathrm{GPU}}/\phi_{\mathrm{GPU}} \approx 0.16$, embodied GPU
carbon adds roughly 16\% on top of operational GPU carbon per TFLOP.

\smallskip
\noindent\textbf{Cumulative lifecycle carbon.}
Total lifecycle carbon combines operational emissions, which scale with
electricity use, and embodied emissions, which are incurred when hardware is
manufactured:
\begin{align}
  \mathcal{F}_{\mathrm{full}}(y)
    &= \underbrace{\frac{\gamma_e}{e}\,E_{\mathrm{full}}^{(g)}(y)}
       _{\mathrm{operational}}
     + \underbrace{\Phi_{\mathrm{HDD}}\sum_{\tau=2025}^{y} A^{(g)}(\tau)}
       _{\mathrm{embodied\ HDD}},
  \label{eq:carbonfull} \\
  \mathcal{F}_{\mathrm{lazir}}(y)
    &= k_{\mathrm{size}}\,\mathcal{F}_{\mathrm{full}}(y) \notag \\
    &\quad + k_{\mathrm{gen}} c\,(\phi_{\mathrm{GPU}} + \Psi_{\mathrm{GPU}})
         \sum_{\tau=2025}^{y}(1-h)\,R^{(g)}(\tau).
  \label{eq:carbon2tier}
\end{align}
The embodied HDD term counts all manufactured drives over the horizon,
including replacements, and therefore scales with the total manufactured
capacity $\sum_\tau A^{(g)}(\tau)$ rather than only the live data volume. In the
two-tier system, both operational and embodied HDD emissions scale by
$k_{\mathrm{size}}$, since the IR store is only a $k_{\mathrm{size}}$ fraction
of the full corpus. Misses add GPU carbon, with embodied GPU emissions
$\Psi_{\mathrm{GPU}}$ charged alongside operational GPU emissions
$\phi_{\mathrm{GPU}}$ per TFLOP.

\begin{table}[t]
  \centering
  \footnotesize
  \setlength{\tabcolsep}{4pt}
  \renewcommand{\arraystretch}{1.1}
  \caption{Cumulative 20-year lifecycle carbon emissions per modality.
           The setting is $g{=}20\%$, $y{=}2045$, and
           $L_{\mathrm{HDD}}{=}5$\,yr. Model-specific
           $(k_{\mathrm{size}},k_{\mathrm{gen}})$ values are taken from
           \autoref{tab:modality-ex}. The full-HDD baseline emits
           $3{,}269{+}1{,}778{=}5{,}047$\,t CO$_2$.}
  \label{tab:carbon}
  \begin{tabular}{@{}lrrrr@{}}
    \toprule
    Scenario & Op.\ (t) & Emb.\ (t) & Total (t) & Saving \\
    \midrule
    Full HDD (baseline)             & 3{,}269 & 1{,}778 & 5{,}047 & --- \\
    \midrule
    LazIR (FLUX.1-dev)   &   908 &   354 & 1{,}262 & 75.0\% \\
    LazIR (MusicGen)     &    16 &     9 &      25 & 99.5\% \\
    \bottomrule
  \end{tabular}
\end{table}

\smallskip
\noindent\textbf{Results by modality.}
Evaluating Equations~\ref{eq:carbonfull} and~\ref{eq:carbon2tier} at the 2045
horizon under $g{=}20\%$ annual ingest growth, using each modality's
$(S,c,k_{\mathrm{size}},k_{\mathrm{gen}})$ from
Appendix~\ref{sec:modality-walkthrough}, yields the lifecycle emissions in
Table~\ref{tab:carbon}. The full-HDD baseline emits 5{,}047\,t CO$_2$ over the
20-year horizon. Of this total, 3{,}269\,t comes from powering the cumulative
corpus, while 1{,}778\,t is embodied in manufactured HDDs, including replacement
drives.

LazIR design substantially reduces emissions across modalities because
both operational and embodied HDD emissions shrink in proportion to the latent
storage fraction $k_{\mathrm{size}}$, while miss decoding contributes only a
small amount of GPU carbon. Image generation remains the largest two-tier case,
emitting 1{,}262\,t CO$_2$, a 75.0\% reduction relative to full HDD storage,
because FLUX.1-dev uses a relatively large latent representation
($k_{\mathrm{size}}{=}1/6$). Music falls to only 25\,t CO$_2$, a 99.5\%
reduction, because its compact representation makes the persisted store a tiny
fraction of the original corpus.

Embodied HDD emissions are material: they account for 35\% of the full-HDD
baseline. Under the two-tier design, these embodied emissions shrink with
$k_{\mathrm{size}}$, so the carbon benefit is largest for modalities whose
intermediate representations are most compact.
\label{sec:appendix_source}
\section{Sources for the hardware trend}
The following link is the source for data points in \autoref{fig:price-trends}.

\begin{table}[H]
  \centering
  \footnotesize
  \setlength{\tabcolsep}{4pt}
  \renewcommand{\arraystretch}{1.12}
  \caption{Archived sources for LTO tape media prices.}
  \label{tab:lto_price_sources}
  \begin{tabular}{@{}ll@{}}
    \toprule
    Year & Generation \\
    \midrule

    2013 & LTO-6 \\
    \multicolumn{2}{@{}p{0.92\linewidth}@{}}{\url{https://web.archive.org/web/20130111023406/http://www.tapeandmedia.com/hp-lto-6-tape-ultrium-tapes.asp}} \\
    \addlinespace[2pt]

    2015 & LTO-7 \\
    \multicolumn{2}{@{}p{0.92\linewidth}@{}}{\url{https://web.archive.org/web/20151213020546/http://www.tapeandmedia.com/hp-lto-7-tape-ultrium-tapes-c7977a.asp}} \\
    \addlinespace[2pt]

    2017 & LTO-8 \\
    \multicolumn{2}{@{}p{0.92\linewidth}@{}}{\url{https://web.archive.org/web/20171226234539/https://www.tapeandmedia.com/lto-8-tape-media-tapes.asp}} \\
    \addlinespace[2pt]

    2021 & LTO-9 \\
    \multicolumn{2}{@{}p{0.92\linewidth}@{}}{\url{https://web.archive.org/web/20210924065625/https://tapeandmedia.com/hpe-lto-9-tape-with-barium-ferrite-bafe-q2079a/}} \\
    \addlinespace[2pt]

    2025 & LTO-10 \\
    \multicolumn{2}{@{}p{0.92\linewidth}@{}}{\url{https://web.archive.org/web/20250621164207/https://tapeandmedia.com/hpe-lto-10-ultrium-tape-30tb-75tb-compressed-rw-data-cartridge-q2080a/}} \\

    \bottomrule
  \end{tabular}
\end{table}

\begin{table}[H]
  \centering
  \footnotesize
  \setlength{\tabcolsep}{4pt}
  \renewcommand{\arraystretch}{1.12}
  \caption{Sources for HDD capacity milestones.}
  \label{tab:hdd_capacity_sources}
  \begin{tabular}{@{}llp{0.68\linewidth}@{}}
    \toprule
    Year & Capacity & Source \\
    \midrule
    2010 & 3 TB  & \url{https://www.storagereview.com/news/hitachi-deskstar-7k3000-deskstar-5k3000-and-xl-external-hard-drives-released} \\
    2012 & 4 TB  & \url{https://www.businesswire.com/news/home/20111212005377/en/Hitachi-GST-Ships-Two-New-4TB-Deskstar-Based-Solutions} \\
    2013 & 6 TB  & \url{https://www.prnewswire.com/news-releases/hgst-ships-6tb-ultrastar-he6-helium-filled-drives-for-high-density-massive-scale-out-data-center-environments-230452991.html} \\
    2014 & 8 TB  & \url{https://www.storagereview.com/review/hgst-ultrastar-helium-he8-8tb-enterprise-hard-drive-review} \\
    2015 & 10 TB & \url{https://www.anandtech.com/show/9830/western-digital-expands-hgst-helium-drive-lineup-with-10tb-ultrastar-he10} \\
    2017 & 12 TB & \url{https://www.techpowerup.com/232748/hgst-ships-ultrastar-he12-12tb-enterprise-hard-drives} \\
    2018 & 14 TB & \url{https://www.anandtech.com/show/11901/western-digital-now-shipping-14tb-hgst-ultrastar-hs14} \\
    2019 & 16 TB & \url{https://investors.seagate.com/news/news-details/2019/Seagate-Delivers-Industrys-First-Enterprise-Ready-Exos-16TB-Hard-Drive-And-Raises-The-Bar-With-IronWolf-16TB-For-NAS/default.aspx} \\
    2020 & 18 TB & \url{https://blocksandfiles.com/2020/08/17/wd-seagate-18tb-disk-drive-race/} \\
    2021 & 20 TB & \url{https://www.servethehome.com/seagate-exos-x20-and-ironwolf-pro-20tb-hdds-shipping/} \\
    2022 & 22 TB & \url{https://www.westerndigital.com/company/newsroom/press-releases/2022/2022-07-19-western-digital-extends-hdd-technology-and-areal-density-leadership} \\
    2023 & 24 TB & \url{https://investors.seagate.com/news/news-details/2023/Seagates-New-Exos-24TB-Hard-Drives-Deliver-Market-Leading-Capacity-for-Hyperscalers-and-Enterprise-Data-Centers/default.aspx} \\
    2024 & 30 TB & \url{https://www.storagenewsletter.com/2024/01/18/seagate-mozaic-3-ww-highest-capacity-hdd-at-30tb-and-shipping-in-1q24/} \\
    2025 & 32 TB & \url{https://www.tomshardware.com/pc-components/hdds/seagate-launches-32tb-exos-m-hard-drive-based-on-hamr-technology-mozaic-3-drives-are-the-worlds-first-generally-available-hamr-hdds} \\
    \bottomrule
  \end{tabular}
\end{table}

\paragraph{GPU price} Found from \url{https://epoch.ai/data/machine-learning-hardware}

\end{document}